\documentstyle[psfig,11pt]{article}
\begin{document}
%
% preprint number
{\hfil\hfill \bf EFI 96-25}
%
%
% title and author list
%
\vspace{3mm}
\begin{center}

{\Large \bf A High Statistic Search for Ultra-High Energy}

\vspace{1mm}

{\Large \bf Gamma-Ray Emission from Cygnus X-3 and Hercules X-1}

\vspace{6mm}

A. Borione, M.C. Chantell, C.E. Covault, J.W. Cronin, B.E. Fick,
J.W. Fowler, L.F. 

Forston, K.G. Gibbs, K.D. Green, B.J. Newport, R.A. Ong,
S. Oser, L.J  Rosenberg$^*$

{\em The Enrico Fermi Institute, The University of Chicago,
Chicago, IL 60637, USA}

\vspace{5mm}

M.A. Catanese$^\dag$, M.A.K. Glasmacher, J. Matthews,
D. Sinclair, J.C. van der Velde

{\em Department of Physics, The University of Michigan,
Ann Arbor, MI 48109, USA}

\vspace{5mm}

D.B. Kieda

{\em Department of Physics, The University of Utah,
Salt Lake City, UT 84112, USA}

\vspace{4mm}

\centerline{(To be published in Physical Review D)}

\vspace{1mm}

\end{center}

%******************************************************************************
% START OF TEXT
%******************************************************************************

\section*{Abstract}
We have carried out a high statistics ($2 \times 10^9$ events)
search for ultra-high energy
gamma-ray emission from the X-ray binary sources Cygnus X-3 and
Hercules X-1.
Using data taken with the CASA-MIA detector over a five year period
(1990-1995), we find no evidence for steady emission from either
source.
The derived 90\% c.l. upper limit to the steady integral flux
of gamma-rays from Cygnus X-3
is $\Phi (E > 115\,{\rm TeV}) 
< 6.3 \times 10^{-15}$ photons cm$^{-2}$ sec$^{-1}$,
and from Hercules X-1 it is $\Phi (E > 115\,{\rm TeV}) 
< 8.5 \times 10^{-15}$
photons cm$^{-2}$ sec$^{-1}$.
These limits are more than two orders of magnitude lower than earlier
claimed detections and are better than recent experiments
operating in the same energy range.
We have also searched for transient emission on time periods of
one day and $0.5\,$hr and find no evidence for such emission from
either source.
The typical daily limit on the integral gamma-ray flux
from Cygnus X-3 or Hercules X-1 is
$\Phi_{{\rm daily}} (E > 115\,{\rm TeV}) < 2.0 \times 10^{-13}$
photons cm$^{-2}$ sec$^{-1}$.
For Cygnus X-3, we see no evidence for emission correlated with the
$4.8\,$hr
X-ray periodicity or with the occurrence of large
radio flares.
Unless one postulates that these sources were very active earlier
and are now dormant,
the limits presented here put into question the earlier results,
and highlight the difficulties that possible future experiments
will have in detecting gamma-ray signals at ultra-high energies.

% Main text starts here

\section{Introduction}
\label{sec:intro}
Cosmic ray particles span a remarkable range of energies, from
the MeV scale to more than $10^{20}\,$eV (eV = electron Volt).
At energies above $1\,$TeV ($10^{12}\,$eV), we know that
cosmic rays do not originate from local sources in or nearby our
Solar System.
Therefore, high energy cosmic rays must come from acceleration sites in the
Galaxy at large or from outside the Galaxy.
Remarkably, after many years of research, the exact sites of
high energy cosmic ray acceleration remain unknown.

There are several difficulties that plague efforts to pinpoint the
origins of high energy cosmic rays.
First, since the bulk of the cosmic radiation is electrically
charged, any source information contained in the directions of the
arriving particles is lost due to deflection in the 
Galactic magnetic field.
A second difficulty concerns the energetics of 
the proposed cosmic ray acceleration
mechanisms.
For example, 
although models based on shock acceleration in supernova remnants offer
plausible explanations for the cosmic ray origin up to $10^{14}\,$eV
(and perhaps up to $10^{15}\,$eV), these models become less
satisfying and less realistic at energies above $10^{15}\,$eV.
Since cosmic ray origins remain
mysterious, it is natural to search for neutral
radiation from point sources 
which, if seen,
could pinpoint possible
acceleration sites.
The question of cosmic ray origin is thus a prime motivation for high
energy neutral particle (gamma-ray or neutrino) astronomy.

In addition to supernova remnants, possible galactic sources of high
energy particles include pulsars and compact binary systems.
Gamma-radiation has been unambiguously detected from the Crab
Nebula (a supernova remnant) at energies up to $10\,$TeV by
ground-based detectors \cite{ref:Snowmass}, but, historically,
the compact binary sources Cygnus X-3 and Hercules X-1 have received
considerably greater attention in the ground-based astronomical community.
In the period 1975-1990, literally dozens of 
gamma-ray detections from Cygnus X-3 and Hercules X-1
were reported by numerous experiments.
The detections spanned
a wide range of energies ($100\,$MeV to $10^{17}\,$eV)
\cite{ref:Reviews},
were generally of low statistical significance
(typically three to four standard deviations), and episodic in nature.
Often a statistically significant signal could only be extracted
as a result of a periodicity analysis, where the
data were phase-locked to a known source X-ray periodicity.
In spite of these difficulties, the sheer number of reports made
it difficult to dismiss the detections as being
entirely due to statistical fluctuation \cite{ref:Protheroe}.
In fact, by the late 1980's, it was generally established that
Cygnus X-3 and Hercules X-1 were powerful
emitters of high energy gamma-rays (although 
contrary interpretations of the data existed \cite{ref:Chardin}).
A number of new, more sensitive
ground-based air shower arrays were commissioned
at this time, including the CASA-MIA experiment in Dugway, Utah (USA).
This paper describes long-term (1990-1995) observations of Cygnus
X-3 and Hercules X-1 by CASA-MIA.

In the following section, we
outline the experimental techniques of gamma-ray astronomy.
We summarize the properties of
Cygnus X-3 and Hercules X-1, and
review previous observations
of these sources and the astrophysical ramifications from the
observations.
We then turn our attention to the experimental apparatus,
the event reconstruction procedures, and the methods used to
select gamma-ray candidate events.
We present results from a 
data sample of $2 \times 10^9$ events, and
conclude with comparisons of our results to those of
earlier and contemporaneous experiments.

\section{Experimental Techniques of Gamma-Ray Astronomy}
\label{sec:techniques}
Gamma-ray sources typically exhibit power law spectra;
fluxes fall rapidly with increasing energy.
Space-borne experiments (on satellites or balloons) 
currently have sufficient
sensitivity to detect gamma-rays up to an energy of $\sim 10\,$GeV.
Astronomy at higher energies requires very large collection
areas available only to ground-based telescopes.
Ground-based instruments rely upon the fact that
high energy gamma-rays interact in the Earth's atmosphere to
produce extensive air showers.
At energies near $1\,$TeV, 
atmospheric Cherenkov telescopes 
use optical techniques to
detect the Cherenkov radiation in
the shower.
At higher energies ($\sim 10\,$TeV and above),
the charged particles in the shower penetrate to ground level.
Here, air shower arrays 
record the arrival times and particle densities of the
charged particles.
At energies above $10^{17}\,$eV, there is enough energy in the shower
to allow the detection of 
nitrogen fluorescence in the atmosphere.
The faint near ultraviolet fluorescence signal can be optically detected at
night by experiments such as the Fly's Eye.

\section{Discussion of the Sources and Earlier Results}
\label{sec:sources}
\subsection{Cygnus X-3}
\label{subsec:cygnus}
Cygnus X-3 is one of the most luminous X-ray sources in our
Galaxy \cite{ref:Giacconi}.
The X-ray emission is characterized by a $4.8\,$hr 
periodicity,
which is assumed to be associated with the orbital motion of
a compact object (neutron star or black hole) around its binary
companion.
The periodicity has been well studied; a complete ephemeris
is available for the period from 1970-1995 
\cite{ref:Parsignault,ref:vanderKlis1,ref:vanderKlis2,ref:Kitamoto}.
In addition to being a powerful X-ray source, Cygnus X-3 is seen
in the infrared and is a strong and variable radio source.
Radio flares have been detected in which the output from the source
increases by two to three orders of magnitude on the time scale of
days 
\cite{ref:Gregory1,ref:Johnston,ref:Waltman1,ref:Waltman2}.
These flares were first detected in 1972 and the
outbursts have continued through 1994.
Since
Cygnus X-3 lies in the galactic plane, its optical emission
is largely obscured by interstellar material.
The lack of a strong optical signal makes determination of
the distance to Cygnus X-3 difficult, but general
consensus places it near $10\,$kpc \cite{ref:Lauque}.

The first published result claiming the detection of
gamma-ray emission from Cygnus X-3 
came in 1977 from the SAS-2 satellite 
at low gamma-ray energies (E$> 35\,$MeV) \cite{ref:Lamb1}.
This result made use of an apparent correlation between the
gamma-ray arrival times and the $4.8\,$hr
X-ray periodicity.
Later observations by the COS-B satellite 
\cite{ref:Bennett}
with a much larger
source exposure failed to confirm the SAS-2 result, and the COS-B
authors argued that the initial detection \cite{ref:Lamb1}
was flawed because of an incorrect treatment of the diffuse
gamma-ray component \cite{ref:Hermsen}.
To complicate matters, there have been re-examinations
of both the SAS-2 \cite{ref:Fichtel} and COS-B \cite{ref:Li}
data sets which claim that 
signals indeed exist in both cases.
Most recently, the EGRET experiment on the Compton Gamma-Ray Observatory
(CGRO) failed to detect gamma-ray emission from Cygnus
X-3 at a sensitivity level comparable to COS-B
\cite{ref:Michelson}.
To summarize,
there exists some controversy as to whether low energy
gamma-rays have {\it ever} been detected from Cygnus X-3.
Regardless, it can be reasonably concluded
that the source is not a strong emitter of gamma-rays in the energy
range between $30\,$MeV and $10\,$GeV.

The first published report of very-high energy
gamma-ray emission from Cygnus X-3
came from the Crimean Observatory 
using an atmospheric Cherenkov telescope at energies above $2\,$TeV 
\cite{ref:Neshpor}.
This result was based on data taken between 1972 and 1977 and 
the emission was claimed to be correlated with the
$4.8\,$hr
X-ray period.
From 1980 to 1990, there were numerous additional detections
of Cygnus X-3 by atmospheric Cherenkov telescopes 
\cite{ref:Danaher,ref:Lamb2,ref:Dowthwaite1,ref:Cawley1,%
ref:Chadwick,ref:Bhat,ref:Brazier}.
The detections were generally episodic in nature and 
usually required the
use of the 
$4.8\,$hr
periodicity to extract a signal.
Evidence for a 12.6 msec gamma-ray pulsar inside the Cygnus X-3 system
was claimed on more than one occasion
\cite{ref:Chadwick,ref:Brazier,ref:Gregory2}.

At the higher energies accessible by ground arrays, 
evidence for ultra-high energy gamma-ray emission from
Cygnus X-3 was presented by the Kiel array \cite{ref:Samorski}
and subsequently by the Haverah Park experiment \cite{ref:Lloyd}.
These results were based on data taken between 1976 and 1980.
The gamma-ray emission was apparently steady over this time
period and was
correlated with the X-ray
periodicity.
Additional evidence  
for gamma-ray emission from Cygnus X-3
was later reported by other air shower detectors
\cite{ref:Kifune,ref:Alexeenko,ref:Baltrusaitis1,%
ref:Tonwar1,ref:Morello}.

At extremely-high energies (E $> 5 \times 10^{17}\,$eV),
evidence was presented for neutral particles from the direction
of Cygnus X-3 by the Fly's Eye \cite{ref:Cassiday1} and
by Akeno \cite{ref:Teshima} groups, based on data taken 
during the periods 1981-1989 and 1984-1989, respectively.
These data apparently indicated steady emission of neutral
particles from Cygnus X-3 that was uncorrelated with the X-ray periodicity.
The Haverah Park experiment, operating in the same energy range, and
during much of the same period in time,
found no evidence for such emission
\cite{ref:Lawrence}.

\begin{figure}
\centerline{\ \psfig{figure=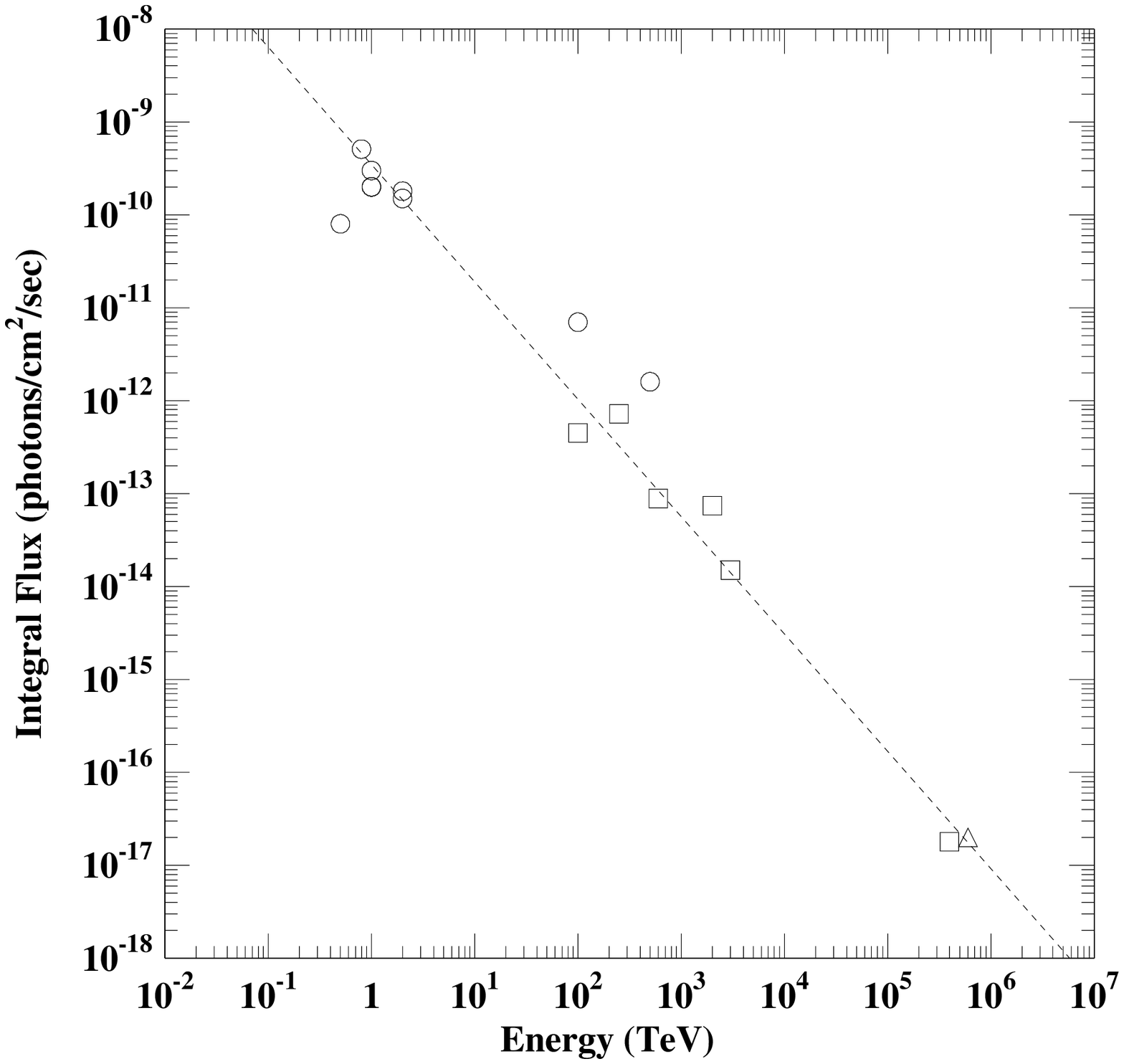,height=17.0cm}}
\caption{
Published results 
from ground-based experiments indicating evidence for gamma-ray
emission from Cygnus X-3 during the period 1975-1990.
The circles indicate results from atmospheric
Cherenkov telescopes, the squares show data taken
by air shower arrays, and the triangle indicates the
result from the Fly's Eye experiment.
The dashed curve is an approximate power law fit to the data with
a slope of $-1.1$.
Not shown in this figure are several upper limits to the flux
of gamma-rays from Cygnus X-3 during this same epoch.
The two points at extremely high energy ($5\times 10^{5}\,$ TeV)
have been slightly displaced from each other
for clarity.}
\label{fig:OldCyg}
\end{figure}

The evidence 
from ground-based experiments 
for gamma-ray emission from Cygnus X-3 from 1975 to 1990
is shown in Figure~\ref{fig:OldCyg}.
Here,
the integral gamma-ray fluxes
are plotted as a function of energy.
Also
shown is a single power law fit of the
form E$^{-1.1}$.
The fact that the gamma-ray fluxes at widely varying energies
could be approximately fit by a single power law was taken by
some as evidence of a unified acceleration mechanism at the source.
One should note, however, that all results
shown in Figure~\ref{fig:OldCyg} represent {\it integral}
flux measurements by experiments incapable of accurately measuring
differential fluxes.
Since the detections were generally
only marginally statistically significant, the reported fluxes
equally represent the three to four standard deviation
sensitivity of each instrument at a fixed energy.
The fluxes
therefore would naturally fall on an E$^{-1}$ power law,
if the sensitivities of the experiments scaled linearly
with energy (which was approximately true for these
first-generation experiments).
It has also been pointed out that even if the source
mechanism produced emission with a single power law form,
the {\it detected} flux at Earth would have a significant
dip between $10^{3}$ and $10^{4}\,$TeV 
because of absorption of gamma-rays by
the cosmic microwave radiation \cite{ref:Cawley2}.

Starting with the CYGNUS experiment in 1988 \cite{ref:Dingus1},
a number of more sensitive ground-based experiments were
unable to detect gamma-ray emission from Cygnus X-3,
at levels significantly lower than the earlier reports.
Upper limits on the flux were reported for experiments using both the
atmospheric Cherenkov technique \cite{ref:Fegan},
as well as the ground-array technique
\cite{ref:Cassiday2,ref:Alexandreas1}.
Using parts of the eventual CASA-MIA detector, some
of us reported limits for data taken in 1988-1989
\cite{ref:Ciampa} and in 1989 \cite{ref:Cronin}.
The general trend of a ``quiet'' Cygnus X-3
continued into the early 1990's, although
there were several publications claiming gamma-ray emission based
largely on data that had been taken in the previous decade
\cite{ref:Muraki,ref:Bowden,ref:Tonwar2}.

\subsection{Hercules X-1}
\label{subsec:hercules}
Like Cygnus X-3, Hercules X-1 is a powerful binary X-ray 
source \cite{ref:Tananbaum}.
The X-ray emission is modulated on a time scale of 1.7 days
which is assumed to result from the
orbital motion of the binary pair.
Unlike Cygnus X-3, Hercules X-1 is not seen in radio, but
has been observed for many years in the optical 
range \cite{ref:Jones} and a $5.8\,$kpc distance
to the source has been determined \cite{ref:Forman}.
In addition, Hercules X-1 contains an X-ray pulsar
with a period of $1.24\,$sec \cite{ref:Tananbaum}, but
whose ephemeris is relatively poorly determined because of
unpredictable variations in the spin-up rate \cite{ref:Deeter}.
Hercules X-1 has not been detected by space-borne gamma-ray instruments.

The first evidence from a ground-based observatory for gamma-ray emission
from Hercules X-1 was reported in 1984 by the Durham group using
the atmospheric Cherenkov technique \cite{ref:Dowthwaite2}.
The reported gamma-ray emission came in the form of a short
burst ($\sim 3$ minute duration) that exhibited 1.24 sec
periodicity.
Following this report, additional pulsed emission was claimed by
Cherenkov detectors operating at ultra-high energies (E $> 500\,$TeV)
\cite{ref:Baltrusaitis2} and at TeV energies
\cite{ref:Gorham1,ref:Gorham2}.

The most intriguing evidence for gamma-ray emission from
Hercules X-1 came from data taken in 1986 by three
experiments.
Data taken between April and July of 1986 by 
the Haleakala \cite{ref:Resvanis}
and Whipple \cite{ref:Lamb3} telescopes operating
near $1\,$TeV, and by the CYGNUS experiment \cite{ref:Dingus2}
in July of 1986 operating at energies above
$50\,$TeV, all indicated evidence for gamma-ray emission from Hercules
X-1 in the form of short bursts of approximately 0.5 hr duration.
In addition, the emission detected by each experiment exhibited
a common periodicity near $1.2358\,$sec,
which differed by a significant amount 
($\sim 0.16\%$ lower)
from the known X-ray period.
The data from the CYGNUS experiment was further puzzling
because the events from the direction of Hercules X-1
had a muon content that was similar to the cosmic ray background events,
whereas gamma-ray events should have contained
significantly fewer muons.
Later, two groups with somewhat poorer sensitivity presented
additional evidence for gamma-ray emission from Hercules X-1 at
different times in 1986 
at TeV \cite{ref:Vishwanath} and $100\,$TeV \cite{ref:Gupta}
energies.

Since the advent of upgraded and improved experiments in the early
1990's, the gamma-ray signals from Hercules X-1 disappeared
from the published literature.
The Whipple group, using a more sensitive Cherenkov imaging
technique, failed to detect emission from Hercules X-1, and found 
no statistically significant evidence for gamma-ray
emission from Hercules X-1 over a six year period, 
even including their data from 1986 \cite{ref:Reynolds}.
An enlarged and improved
CYGNUS experiment also failed to see gamma-rays from
Hercules X-1 
in the period between 1987 and 1991
\cite{ref:Alexandreas2}.
Using data taken in 1989,
we reported upper limits on the emission of gamma-rays from
Hercules X-1 using part of the eventual CASA-MIA experiment \cite{ref:Cronin}.

\subsection{Theoretical Implications}
\label{subsec:theory}
The many claims of very high energy gamma-ray
emission from the binary systems Cygnus X-3 and Hercules
X-1 fueled great interest in the development of astrophysical
models to explain such emission.
There were also non-standard particle physics models put forward
to explain the observations; these models will not be discussed here.
 
For the case of Cygnus X-3, where the gamma-ray emission was
generally observed with a 
$4.8\,$hr
periodicity, the astrophysical
models needed to incorporate the orbital dynamics of the
binary system.
Models in which an energetic pulsar alone served as the power source
for the gamma-rays \cite{ref:Cheng1}
or in which accretion powered the gamma-rays
\cite{ref:Chanmugam}
were proposed.
These models generally had
difficulty
in producing gamma-rays at energies above $10^{15}\,$eV.
More popular were a 
general class of models in which the gamma-rays were produced from
the decays of $\pi^0$ mesons made in hadronic collisions
\cite{ref:Vestrand,ref:Eichler1,ref:Kazanas}.
The hadronic beam 
resulted from
diffusive shock acceleration,
perhaps near the neutron star, and possible beam targets included
the
atmosphere of the companion star or material in the accretion disk.
Such ``beam-dump'' models were capable of explaining
gamma-rays at ultra-high
energies and were also able to accommodate the observed periodicities of
the gamma-ray signals.
Several authors recognized
that in order to explain
the ultra-high energy gamma-ray fluxes initially seen,
the required luminosity of Cygnus X-3 
would also be sufficient to account for
a substantial fraction of the high energy cosmic ray flux
\cite {ref:Wdowczyk}.
Hillas pointed out that if Cygnus X-3 consisted of a $10^{17}\,$eV
accelerator with a luminosity of $\sim 10^{39}\,$ergs/sec,
only one such object like it would be required to explain the origin
of cosmic rays above $10^{16}\,$eV \cite{ref:Hillas}.

Unlike Cygnus X-3, the gamma-ray emission from Hercules X-1 was not
seen to be correlated with the orbital motion of the binary system,
but instead with the pulsar periodicity.
This observation, along with evidence that the emission appeared in
the form of short bursts,
led naturally to models in which the pulsar itself was the 
power source.
In such models,
the gamma-rays were produced by the interaction
of a charged particle beam with the accretion disk
\cite{ref:Eichler2,ref:Gorham3}.
More difficult to explain were the 1986 observations of
gamma-ray emission at a slightly shorter period than the X-ray period.
The anomalous gamma-ray periodicity was
explained by the presence of matter in the accretion disk
which periodically obscured the gamma-ray interaction region
\cite{ref:Cheng2,ref:Slane}.

In summary, although theoretical difficulties existed
in explaining the apparent signals of gamma-rays from
Cygnus X-3 and Hercules X-1, the signals were
tantalizing because of the possibility that they
revealed important sources of the ultra-high
energy cosmic ray flux.

\section{Experimental Procedure}
\label{sec:experiment}
\subsection{The CASA-MIA Experiment}
\label{subsec:casamia}
The CASA-MIA experiment is located in Dugway, Utah, USA 
($40.2^\circ\,$N, $112.8^\circ\,$W) at an altitude
of $1450\,$m above sea level ($870\,$g/cm$^2$ atmospheric depth).
CASA-MIA consists of two major components: the Chicago
Air Shower Array (CASA), a large surface array of scintillation detectors,
and the Michigan Array (MIA), a buried array of scintillation counters
sensitive to the muonic component of air showers.

CASA consists of 1089 scintillation detectors placed on a 
$15\,$m square grid and enclosing an area of $230,400\,$m$^2$.
Construction on the array started in 1988 and a small portion
(5\%) of the experiment operated in 1989. A more
substantial portion ($\sim$50\%) of it was completed by early 1990.
Data collection with this portion 
started on March 1, 1990.
Additional detectors were added in 1990 
to complete the construction.

MIA consists of 1024 scintillation counters located beneath CASA in
16 groupings (patches).
The total active scintillator area is $2,500\,$m$^2$ and the counters
are buried beneath $\sim 3.5\,$m of soil.
This depth corresponds to a muon threshold energy of approximately $0.8\,$GeV.
Parts of MIA were operational as early as 1987, 50\% of the experiment
was completed by early 1990, and the entire array was working
by early 1991.
The CASA-MIA
experiment was turned off temporarily in 1991 for repair due to
lightning damage, but has operated essentially uninterrupted since
that time.
Table~\ref{tab:array} summarizes the size and detector makeup of 
CASA-MIA as a function of time.

\begin{table}
\begin{center}
\caption{
Size and makeup of CASA-MIA experiment as a function of time
for the data sample used in this analysis.
A range of values indicates that the experiment was
being enlarged during this period of time.
Data taken after August 1995 are not used in this analysis.}
\label{tab:array}
\vspace{10pt}
\begin{tabular}{|cccc|}\hline
Epoch & Enclosed Area (m$^2$) & CASA Detectors & MIA Counters \\
\hline
Mar. 1990 -- Oct. 1990 &         108,900 &      529 & 512 \\
Oct. 1990 -- Apr. 1991 & 108,900--230,400 & 529--1089 & 512--1024 \\
Jan. 1992 -- Aug. 1993 &         230,400 &     1089 &     1024 \\
Aug. 1993 -- Aug. 1995 &         216,225 &     1056 &     1024 \\
\hline
\end{tabular}
\end{center}
\end{table}

\begin{figure} 
\centerline{\ \psfig{figure=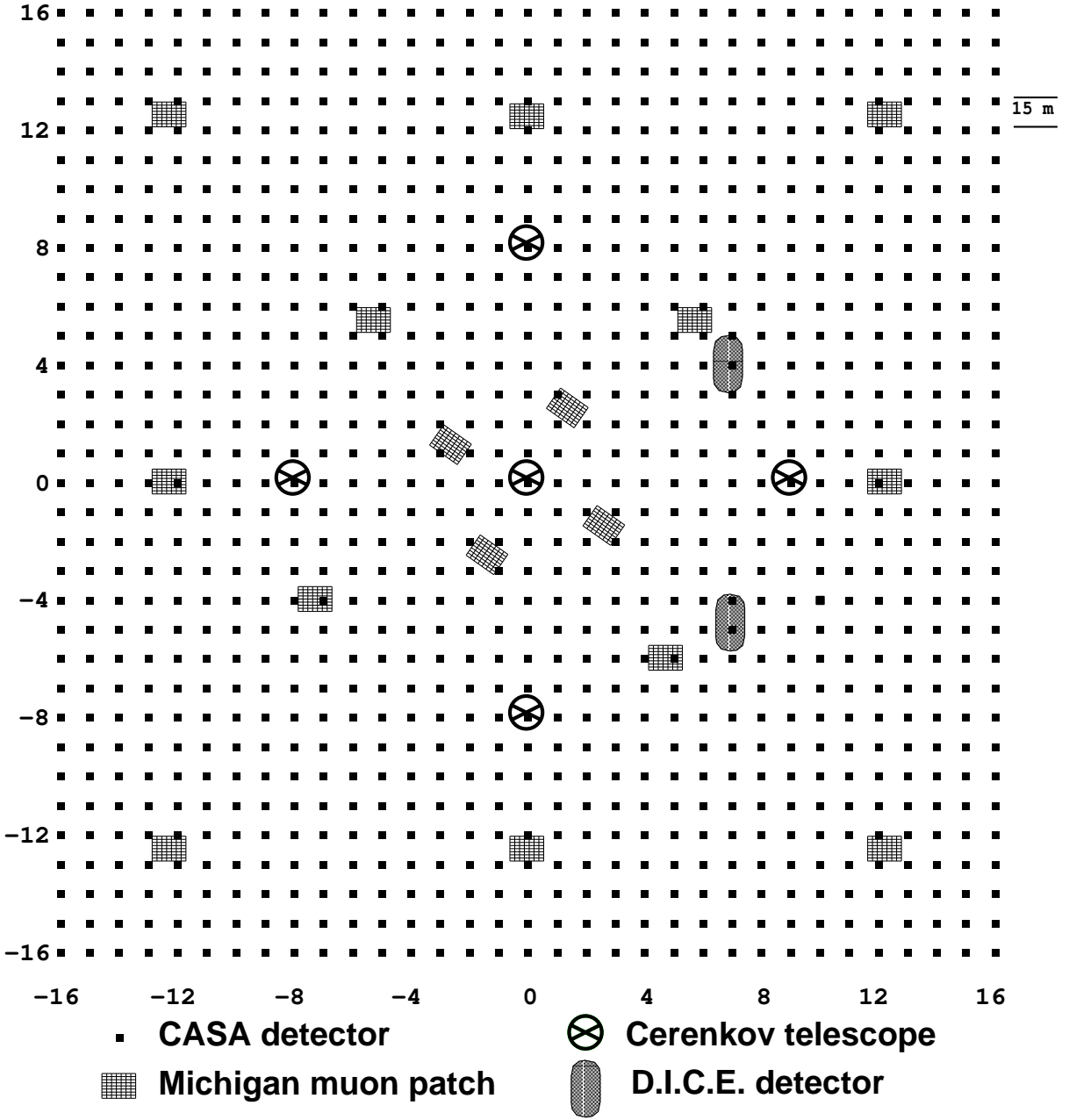,height=17.0cm}}
\caption{
Plan view of the CASA-MIA experiment (Dugway, Utah, USA).
Small squares indicate the 1089 scintillation detectors of
the CASA surface array.
Sixteen large rectangles indicate the patches of
scintillation counters of the
MIA underground array (64 counters/patch).
Five crossed circles indicate tracking
Cherenkov telescopes.
The D.I.C.E. detectors are not used
in this analysis.}
\label{fig:array}
\end{figure}

Figure~\ref{fig:array} shows a plan view of the experimental site.
In addition to CASA-MIA, there are other installations
at the same site.
The other equipment used in this analysis is an array
of five tracking Cherenkov telescopes.
One telescope is located at the center of CASA-MIA, and the other
four are 120$\,$m away from the center along the major axes.
Each telescope consists of a $35\,$cm diameter mirror which focuses
Cherenkov radiation onto a single $5.1\,$cm photomultiplier tube (PMT).
The signals from the PMTs are digitized to record the amplitude and
time of arrival of the Cherenkov wavefront at each telescope location.
The shower direction is reconstructed
by fitting the Cherenkov arrival times to a conical wavefront.

A complete description of the CASA-MIA experiment can be found
elsewhere \cite{ref:NIMPaper}; here we briefly describe some aspects
of the experiment that are relevant for this analysis.
Each CASA station consists of four scintillation counters connected
to a local electronics board.
A station is {\it alerted} when at least two of the four counters
fire within a $30\,$nsec window and it is {\it triggered} when at least
three of the counters fire.
If three or more stations trigger within a time period of approximately
$3\,\mu$sec, an {\it array trigger} is said to have occurred.
The array trigger rate depends on operating conditions
(e.g. atmospheric pressure), but is
$\sim$20 Hz for the full CASA-MIA experiment.

Upon an array trigger, the Universal Time (UT) is latched and
recorded by either a GOES Satellite Receiver Clock (1990-1993)
with a precision of $\pm 1\,$msec, or by a Global Positioning System (GPS)
clock (1993-1995) with a precision of $\pm 100\,$nsec.
For each array trigger,
a command is broadcast to the array
instructing each alerted CASA station to digitize
and record its data, and
a signal is generated to stop time-to-digital
converters (TDCs) on each MIA counter.
The TDCs have a range of $4\,\mu$sec and 
a least bit precision of $4\,$nsec.
The data from each CASA station consist of the arrival times and pulse-height
amplitudes of the pulses from each scintillation counter, as well
as the arrival times of pulses from the four nearest neighbor stations.
The data from each MIA counter consist of the 
time of arrival of the array trigger
relative to the passage of a muon through the counter.
The CASA-MIA data and the Universal Time recorded as the result
of an array trigger
correspond to a single air shower {\it event}.

\subsection{Event Reconstruction}
\label{sec:recon}
We briefly summarize some of the important aspects of the
CASA-MIA event reconstruction; full details can be found
elsewhere \cite{ref:NIMPaper}.
The data from the experiment are accumulated in runs of six hours
duration.
All calibrations and offsets are 
determined for each run separately.
At the start of a run, the timing constants associated with the CASA
station electronics are calibrated by an internal oscillator.
Timing constants are corrected for the effects of temperature
by studying the constants over the span of a week.
The CASA counter particle gains are determined for each run from
the abundant cosmic ray air showers.
The counter gains are found from the PMT amplitude distributions
of those counters hit in stations with two out of four counters hit.
A statistical correction of $\sim 20\%$
accounts for the fact that
on average slightly more than one particle passes through a counter
in this situation.
The CASA cable and electronic delays 
are determined from the zenith angle distributions of the detected
events.
The relative delay between the CASA and MIA trigger systems is
determined by centering the peak of the muon arrival time distribution
relative to the position of the CASA trigger time.

We estimate the shower
{\it core position} 
by the location on the ground with the highest particle density.
The total number of particles in the shower, or 
{\it shower size}, is determined
by fitting the density samples obtained from the CASA stations to
a lateral distribution of fixed form.
The mean number of alerted CASA stations is 19 and the
mean shower size is $\sim 25,000$ equivalent minimum ionizing
particles.

The {\it shower direction} is determined from the timing information
recorded by CASA.
The relative times between pairs of adjacent alerted CASA stations
are determined.
Each relative time gives a measure of the shower
direction along one axis of the experiment.
The times are weighted by an empirical function of the
local particle density and distance to the shower core, and
are fit to a wavefront which accounts for
the conical shape of the shower front.
The cone slope is approximately $0.07\,$nsec/m.
The shower direction in local coordinates is defined by two
angles.
The zenith angle, $\theta$, is measured with respect to the vertical
direction and the azimuthal angle, $\phi$, is measured with respect
to East in a counter-clockwise manner.

In order to be confident of any astronomical results, it is essential
to measure the  {\it angular resolution} of the experiment.
The resolution has two parts.
The statistical part largely derives from
the intrinsic 
fluctuations in the arrival times of the
shower particles
and from the timing resolution of the CASA counters and electronics.
The systematic contribution derives primarily from the
accuracies of the experiment survey and of
the calculation of timing delays and offsets.

\begin{figure} 
\centerline{\ \psfig{figure=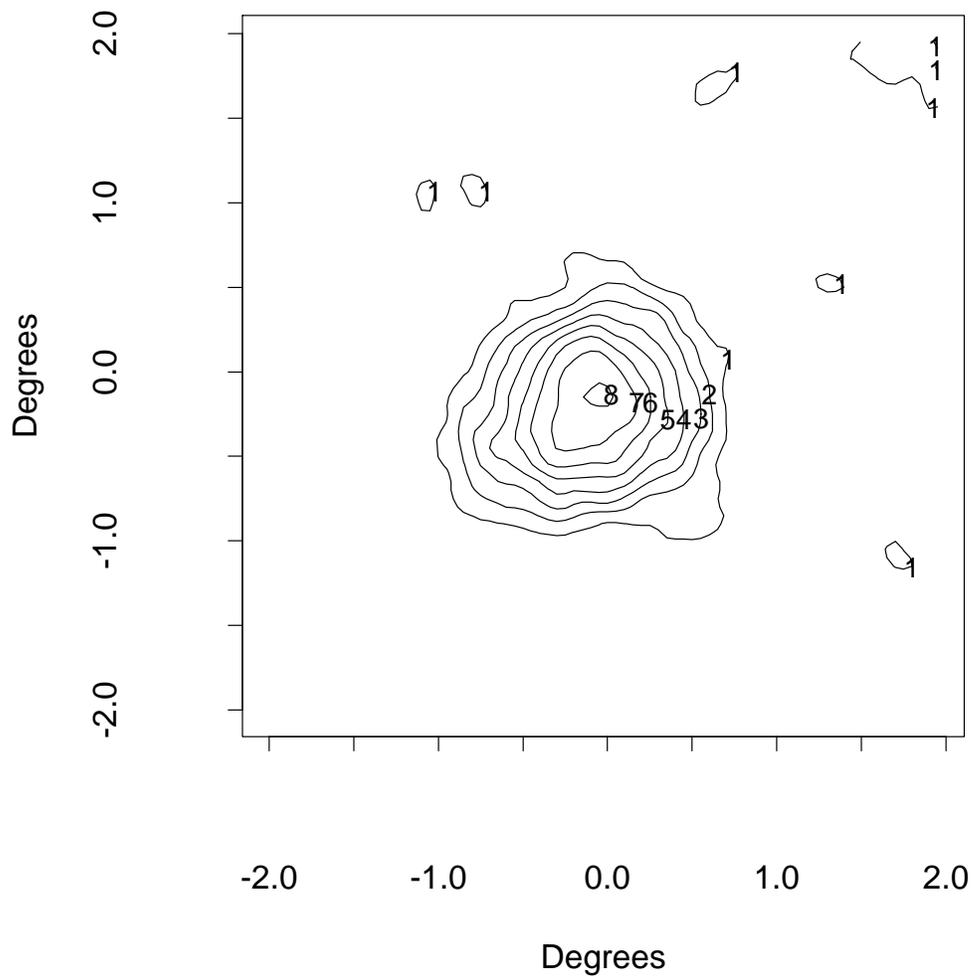,height=17.0cm}}
\caption{
Shadow of the Moon as detected by CASA-MIA.
The contour plot shows the deficit in the
number of detected cosmic ray events 
as a function of the angle from the Moon center.
The axes are defined by the equatorial coordinates of the Moon,
with right ascension along the horizontal axis and
declination along the vertical axis.
The contours correspond to successive one standard deviation
steps.
A Gaussian smoothing factor of $0.5^\circ$ has been applied to the
data.}
\label{fig:moon}
\end{figure}

\begin{figure} 
\centerline{\ \psfig{figure=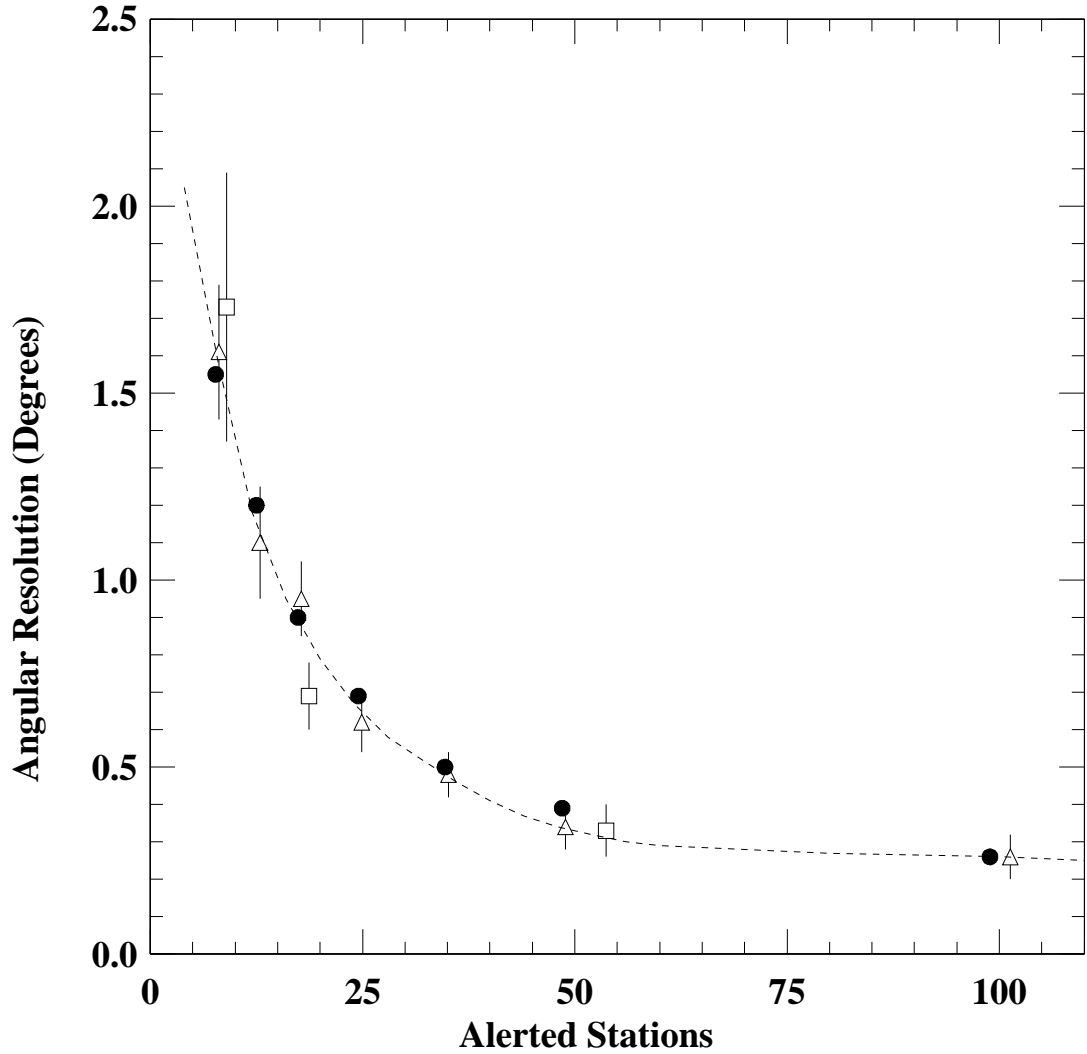,height=17.0cm}}
\caption{
Angular resolution of CASA-MIA as a function of the number
of alerted CASA stations.
The resolution has been estimated by three different
techniques: split-array method (solid points), 
Cherenkov telescope array coincident events (open triangles), and
Moon shadow (open squares).
The dashed curve indicates a simple parametric fit to the
data.}
\label{fig:resolution}
\end{figure}

The statistical contribution to the angular resolution
is determined by three different techniques.
First, on an event-by-event basis, we divide the array into
two overlapping sub-arrays and compare the
shower directions that are reconstructed by each sub-array.
Using an air shower and detector simulation, we estimate
the statistical correction required to derive the angular
resolution from the sub-array direction comparison.
Second, we compare the shower direction as determined by CASA with
the direction determined by the five Cherenkov telescopes for
events in which both CASA and the telescopes triggered.
By statistically removing the angular resolution of the telescopes
from this comparison, we estimate the CASA resolution.
Third, we have detected the shadow that the Moon casts in the cosmic
rays \cite{ref:Moon}.  
For data taken between 1990 and 1995, the Moon 
shadow is shown in Figure~\ref{fig:moon}.
By deconvolving the size of the Moon from the measured shadow,
we obtain another estimate of the angular resolution.
Figure~\ref{fig:resolution} shows the resolution estimates
from these different techniques.
The agreement between the various methods is good, which allows
us to determine a single parametric form for the resolution as
a function of the number of alerted CASA stations.

The systematic contribution to the angular resolution 
of the experiment has been checked by two different
techniques.
First, for data taken in coincidence with the tracking
Cherenkov telescopes, we examine 
the angular difference between the directions determined by CASA
and by the telescopes.
The distribution of these differences indicates
that the systematic offset between
CASA and the telescope array is very small ($< 0.1^\circ$).
The alignment of the telescope array has been verified by the
observation of a number of stars.
A second check on the pointing accuracy of CASA comes from the
Moon shadow.  The center of the Moon shadow image is within 
$0.1^\circ$ of the known position of the Moon.
We conclude that
the pointing uncertainty of CASA is negligible
in comparison with the experiment's angular resolution.

The {\it muon content} of the shower is determined from the
data recorded by MIA.
Since MIA records 
the times of counter hits over an 
interval of $4\,\mu$sec, it is sensitive to muons
produced by showers arriving at any location of the
array and from any direction.
During the same time interval, MIA
also records accidental counter hits produced by PMT noise and by
natural radioactivity in the ground.
The average number of accidental hits is approximately 
sixteen per event,
while
the average number of real muons associated with air showers 
is approximately nine per event.

Real muons arrive within 100$\,$nsec of the shower front arrival, while
the accidental hits occur randomly over the $4\,\mu$sec interval.
We greatly reduce the acceptance for accidental muon hits by narrowing the
time window for accepting muons.
The width of the window is determined from the distribution of
muon times for each six hour run.
We set the width to encompass 95\% of the real shower muons;
on average, it is $\sim 150\,$nsec.
The position of the window is found on an event-by-event basis by
means of a clustering algorithm.
The algorithm searches for the cluster of three or more muons within an
interval of $40\,$nsec.
In approximately 25\% of the events, no cluster is found and the
window position is placed at the center of the muon time distribution
as determined for the entire run.
As a result of tightening the time window for muon hit acceptance,
the average number of real muons recorded is 8.5 per event,
while the average number of accidentals is 0.63 per event.

The CASA-MIA data undergo several stages of processing and compression.
In the most highly compressed format upon which this analysis is based,
the data records are 26 Bytes per event and
include the following information for each event:
Universal Time (UT), number of alerted CASA stations, number of in-time
muon hits, core location, arrival direction, shower size,
and muon shower size (not used here).

\section{Analysis}
\label{sec:anal}
\subsection{Data Sample}
\label{subsec:sample}
The data used in this analysis were taken between March 4, 1990 and
August 10, 1995, with a gap of 255 days in 1991.
The experiment had usable data on 1627 days
with the remainder of the days lost 
largely because of power
outages at the site and computer problems.
The experiment has an instrumental deadtime of approximately
5.4\% which is due to a number of effects, including 
data acquisition computer latency and
the time needed to digitize the CASA station data.
Calibration runs of approximate length of six minutes taken
at the start of data runs,
losses due to $8\,$mm tape failures,
and downtime from array maintenance led
to an additional reduction in the live time to a total of 
1378.4 days (84.7\% of the total).
After the reduction and processing of the data, the final data sample
consisted of $2.0878 \times 10^9$ reconstructed events.

The size of the CASA-MIA data sample is unprecedented in air shower physics.
To ensure data integrity, we
impose a comprehensive
set of data quality cuts.
The cuts are tailored separately for the data sample in which we
only use information from the surface array 
({\it all-data}) and the sample in which we use information from both the
surface and muon arrays ({\it muon-data}).
For each of these samples, we make quality cuts on an event-by-event basis
and on a run-by-run basis.
Cuts are applied to runs and events only in the cases
there there is evidence of an instrumental bias.
The efficiencies of the 
cuts are summarized in Table~\ref{tab:events}.

\begin{table}
\begin{center}
\caption{Quality cut efficiencies and event totals for 
the CASA-MIA data sample.
The muon-data quality cuts are applied after the all-data quality
cuts.
The data sets (all and muon) are described in the text.}
\label{tab:events}
\vspace{10pt}
\begin{tabular}{|ccc|}\hline
  Category &  All-Data Sample  &  Muon-Data Sample \\
\hline
Initial Event Total  &  2087.8M  & 1925.8M \\
Run Cut Efficiency   &   0.935   &  0.929  \\
Event Cut Efficiency &   0.986   &  0.896  \\
\hline
Overall Efficiency   &   0.922   &  0.832  \\
\hline
Final Event Total    &  1925.8M  & 1602.7M \\
\hline
\end{tabular}
\end{center}
\end{table}

For the all-data sample, the run and event cuts have a combined efficiency
of 92.2\%, which yields a final sample of $1.9258\times 10^9$ events.
The most restrictive run cut requires
a minimum fraction of the CASA stations to be working reliably and
removes 2.2\% of the data, largely because of instances in which
isolated parts of the array failed.
For the muon-data sample, the run cuts have an efficiency of 92.9\%.
A cut which requires a sufficient fraction of the muon
counters to be working removes 4.8\% of the data.
The event cuts have an additional efficiency of
89.6\%.
The most restrictive event cut eliminates 3.2\% of the events
because they have no muon information due to deadtime of the
MIA data acquisition system.
The overall efficiency of the muon-data cuts is 83.2\%, which yields
a final sample of $1.6027\times 10^9$ events.

\subsection{Gamma-Ray Selection}
\label{subsec:select}
From prior observations of Cygnus X-3 and Hercules X-1, 
we expect that gamma-ray
fluxes, if present, will be
small in comparison with the isotropic cosmic ray flux.
Therefore, we need to enhance the presence of a possible gamma-ray signal
by eliminating as many cosmic ray air showers as possible, while
keeping a high fraction of the gamma-ray air showers.
To do this, we select
those showers with a reconstructed direction
consistent with the position of the sources (within the angular resolution
of the experiment) and with a muon content
consistent with that expected from a gamma-ray primary.

\subsubsection{Angular Search Bin}
\label{subsec:search}
We define a circular search bin whose size is based on the estimated angular
resolution of the experiment.
For a sufficiently large number of events, the
bin which optimizes the signal-to-noise has a size equal to
$1.59$ times the angular resolution and contains 72\% of the signal.
The CASA-MIA angular resolution depends on
the number of alerted CASA stations in an event, and therefore we use
a variable-sized search bin which scales with the number of alerts.
For simplicity, we use seven different bin sizes that range from
$2.45^\circ$ radius for showers with the least number of alerts,
to $0.41^\circ$ radius for showers with the largest number of alerts.
These bin sizes are shown in Table~\ref{tab:bins}, along with the
fraction of events in each alert range.

\begin{table}
\begin{center}
\caption{Angular search bin sizes and event fractions 
as a function of the number of CASA alerts.
The search bin is a circular region in equatorial coordinates whose
radius is equal to $1.59$ times the angular resolution.}
\label{tab:bins}
\vspace{10pt}
\begin{tabular}{|ccc|}\hline
Alert Range & Event Fraction & Search Bin Radius \\
\hline
{\ 3 - 10 } & 0.331 & $2.45^\circ$ \\
{ 11 - 15 } & 0.224 & $1.88^\circ$ \\
{ 16 - 20 } & 0.121 & $1.40^\circ$ \\
{ 21 - 30 } & 0.150 & $1.05^\circ$ \\
{ 31 - 40 } & 0.064 & $0.78^\circ$ \\
{ 41 - 60 } & 0.058 & $0.60^\circ$ \\
{ $>60$   } & 0.052 & $0.41^\circ$ \\
\hline
\end{tabular}
\end{center}
\end{table}

\subsubsection{Muon Content}
\label{subsec:muon}
Air showers created by gamma-ray primaries are expected to
contain far fewer muons than showers initiated by cosmic ray
nuclei.
This expectation results because the
cross section for
photo-pion production
is much smaller than the cross section for electron-positron
pair production \cite{ref:HERA}.
Therefore,
the interaction of a high energy gamma-ray in the atmosphere
is much more likely to produce an electromagnetic cascade
in the atmosphere than it is to create a hadronic cascade.
Conversely,
cosmic ray nuclei preferentially
interact to create hadronic cascades.
Showers initiated by gamma-rays are thus expected to
contain far fewer
hadrons than those initiated by cosmic rays.
Since air shower muons are predominantly produced from the
decays of pions and kaons in the hadronic cascade,
gamma-ray air showers should contain far fewer muons as well.
Simulations have been done to estimate the muon content of
air showers
\cite{ref:Chatelet,ref:Halzen}.
Our own simulation indicates that an air shower initiated by a 
$100\,$TeV gamma-ray contains, on average, 3-4\% of the number of muons
in a shower initiated by a proton of the same energy.

\begin{figure} 
\centerline{\ \psfig{figure=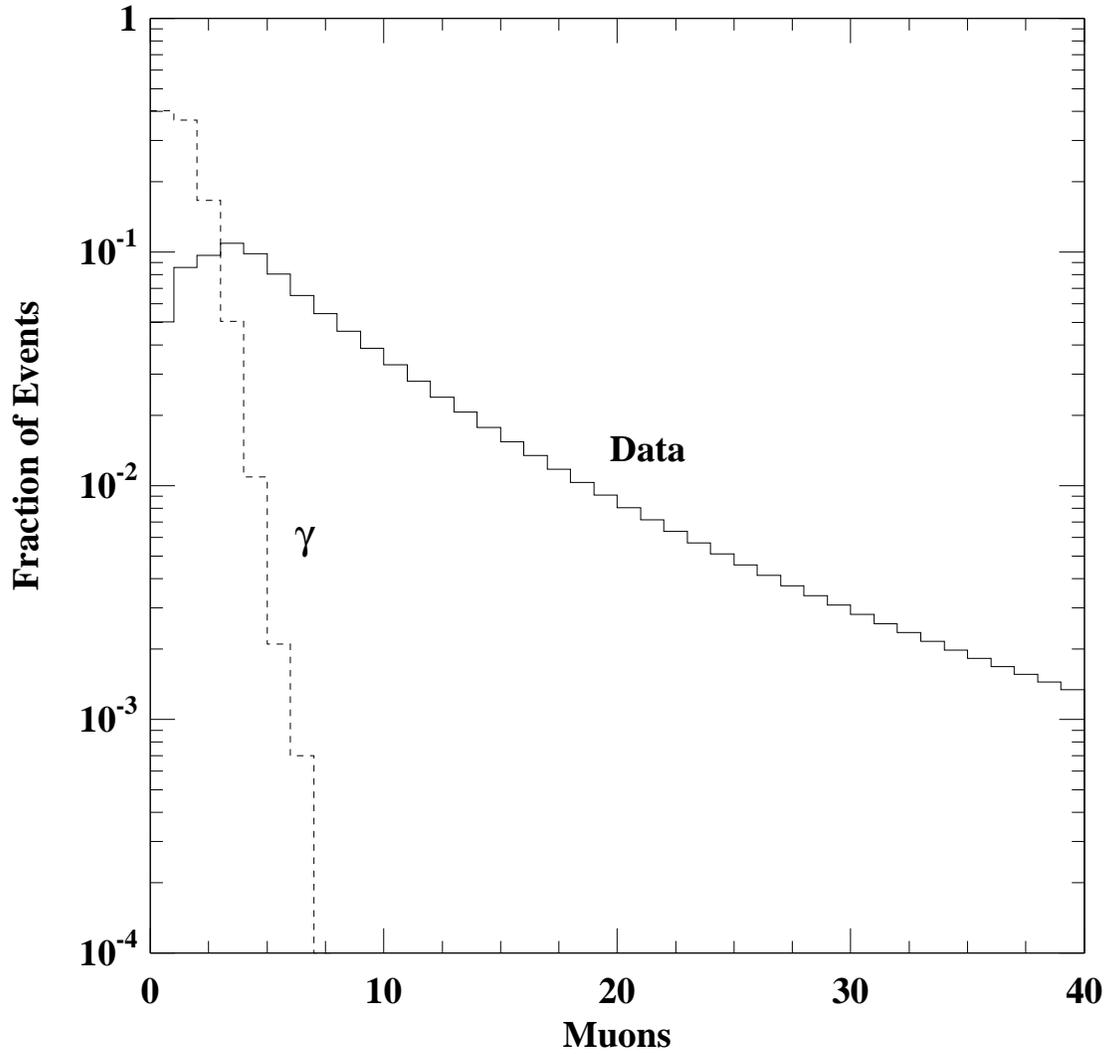,height=17.0cm}}
\caption{
Histograms of the number of muons per event detected by
CASA-MIA for cosmic ray data events (solid)
and simulated gamma-ray events (dashed).
The gamma-ray simulation uses an input energy spectrum with the
same power law spectral index as the cosmic ray 
data.}
\label{fig:in_time_muons}
\end{figure}

The muon content of showers should in principle
be a powerful tool in rejecting cosmic ray background events.
In our experiment, the rejection capability is limited
by the collection area of the muon array and, to a lesser
extent, by the presence of a small amount of accidental
muon hits.
The muon array (MIA) is significantly larger than any other
air shower muon detector built to date, 
but its active area still corresponds to only
$\sim 1\%$ of the enclosed area of the experiment.
As shown in Figure~\ref{fig:in_time_muons}, the average 
of the distribution of the number of in-time muons
is $\sim$8.5, but the shape of the distribution
is such that its mode is three, and a substantial fraction of events
have zero muons.
In Figure~\ref{fig:in_time_muons}, we also show the estimated number of
muons for showers initiated by gamma-rays, including the contribution
from accidental
muon hits.
For gamma-ray showers, we expect, on average, 0.28 real muons per event
and 0.63 accidental muons per event.

In order to enhance a possible gamma-ray signal,
we wish to select {\it muon-poor} events, i.e. events that
have fewer muons than the average expected number.
To do this, we make the assumption that any gamma-ray
signal in the data is much smaller than the flux of cosmic rays.
We can therefore use
the muon information from the detected events
to describe the muon content of the background, and 
our simulation to describe the muon content of the gamma-ray
signal.

The number of muons in a shower depends on a number of observable
quantities, for example,
the number of alerted CASA
stations, shower zenith angle, and core position.
We develop a parameterization for the average number of muons
as a function of these quantities
by examining a large ensemble of actual showers.
We then determine the relative muon content of a specific shower
by comparing the observed muon number, $({\rm n}_\mu)_{{\rm obs}}$,
to the expected number of muons, $<{\rm n}_\mu>_{{\rm exp}}$,
for showers
having similar zenith angles, core positions, and numbers of alerts.
The relative muon content, ${\rm r}_\mu$, is
defined by:

\begin{equation} 
{\rm r}_\mu \ \ \equiv\ \ {\rm Log_{\rm 10}} \
\Biggl[
{  {  ({\rm n}_\mu)_{\rm obs} } \over
   {  <{\rm n}_\mu>_{\rm exp} }        }
\Biggr] \ .  
\label{eq:rmu}
\end{equation}

\begin{figure} 
\centerline{\ \psfig{figure=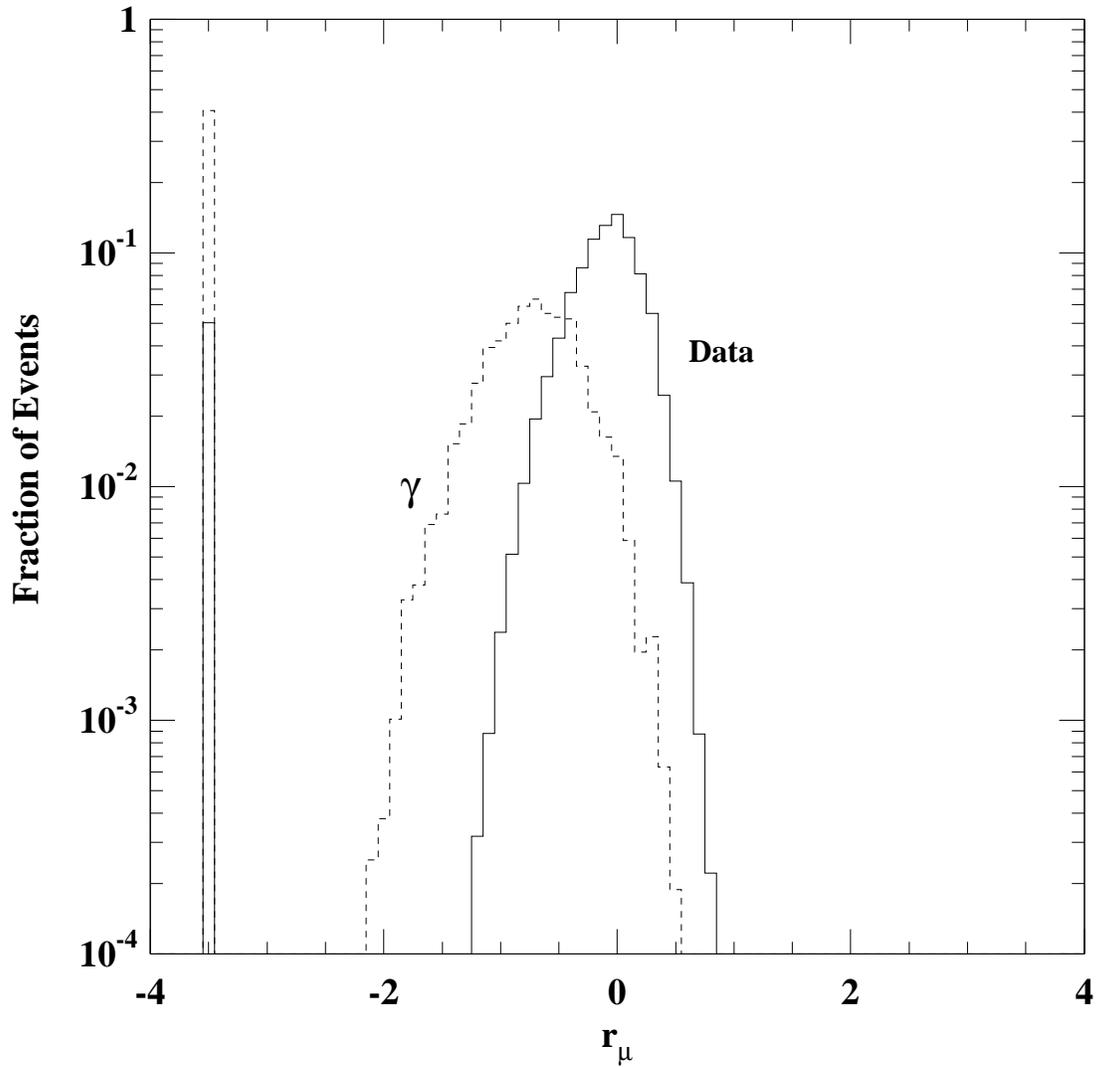,height=17.0cm}}
\caption{
Distributions of the relative muon content, ${\rm r}_\mu$,
for cosmic ray data events (solid) and simulated gamma-ray events
(dashed).
The quantity ${\rm r}_\mu$ is defined in the text. 
Events with zero muons are assigned a value
of ${\rm r}_\mu = -3.5$.
The gamma-ray simulation uses an input energy spectrum with the
same power law spectral index as the cosmic ray 
data.}
\label{fig:rmu}
\end{figure}

\noindent Figure~\ref{fig:rmu} shows the 
distributions of ${\rm r}_\mu$ for 
observed events and for simulated
gamma-ray events.
Muon-poor events are defined as those having 
${\rm r}_\mu$ values less than some cut value.
The position of the cut is chosen to reject as many background
events as possible, while keeping a high fraction of the gamma-ray
events.
The cut value depends weakly on the number of CASA alerts because
the separation between the signal and background 
${\rm r}_\mu$ distributions improves
as the showers get larger.

Table~\ref{tab:rmucut}
shows the ${\rm r}_\mu$ cut values for various samples of data
along with the fractions of signal and background events retained,
and the sensitivity improvement achieved from making a cut.
For the entire data set, the sensitivity is improved by a factor
of 2.94 by cutting on the shower muon content.
The quality factor increases to 29.7 for events having more than
80 alerted CASA stations.

\begin{table}
\begin{center}
\caption{
Quantities associated with the selection of muon-poor events.
Muon-poor events are those having a relative muon content,
${\rm r}_\mu$ (defined in the text),
less than a cut value.
The cut values
are given in the
second column, and
the third and fourth columns give the efficiencies for
passing the cut for gamma-ray signal events and
for hadronic background events, respectively.
The fifth column gives the quality factor, Q, or the
improvement in flux sensitivity from making the cut.}
\label{tab:rmucut}
\vspace{10pt}
\begin{tabular}{|ccccc|}\hline
Data Set    &  ${\rm r}_\mu$ cut Value & Signal $\epsilon$ &
Background $\epsilon$ & Q \\
\hline
All  &  $-0.75$  & 0.72              & 0.0600 & 2.94 \cr
$\le 10$ Alerts &  $-0.50$  &  0.69  & 0.1644 & 1.70 \cr
$ > 10$  Alerts &  $-0.75$  &  0.76  & 0.0362 & 3.99 \cr
$ > 40$  Alerts &  $-1.00$  &  0.71  & $1.77\times 10^{-3}$  &  16.9\cr
$ > 80$  Alerts &  $-1.00$  &  0.77  & $0.67\times 10^{-3}$ & 29.7 \cr
\hline
\end{tabular}
\end{center}
\end{table}

\subsection{Background Estimation}
\label{subsec:back}
We select gamma-ray candidate events 
({\it on-source} events)
based on their
reconstructed arrival direction 
in equatorial coordinates (right ascension, $\alpha$, and 
declination, $\delta$) and on their muon content.
In order to derive the significance of a possible gamma-ray signal,
we need to determine the expected number of background cosmic ray
events 
({\it off-source} events)
that would arrive from the same direction in the sky as the source
and would have a similar muon content as gamma-ray events.
Again,
we make the assumption that the detected air showers
are predominantly caused by background cosmic ray events.
We thus use the detected events themselves to estimate the
expected background.

A common method to estimate the expected number of background events
is to use off-source bins having the same declination as the source,
but having different right ascension values.
This method, which assumes a uniform experiment
exposure over declination,
was satisfactory for earlier smaller experiments.
However, given our
large event sample, 
this technique is not reliable for CASA-MIA
because of small, but non-negligible, systematic biases
(e.g. diurnal variations).
For the CASA-MIA data sample,
a source at a declination of $40^\circ$
occupies an angular bin with
$\sim 1.8 \times 10^6$ events.
The fractional statistical uncertainty corresponding to 
one standard deviation in the number of events is
0.075\%.
In order to accurately estimate the number
of background events, the relative systematic uncertainty must
be well below this level.
As a result, an accurate and robust way to determine
the expected background is needed.
Several methods have been developed 
by other groups 
\cite{ref:Cassiday1,ref:Alexandreas3}
and the method that we use 
is similar to these.

The detection rate of an air shower array 
triggering on cosmic rays 
is determined by the properties of the cosmic ray flux
and by the properties
of the array itself.
Assuming that the cosmic ray parameters do not change with time,
any variation in the detection rate is caused only by
changes in the detector or in the atmospheric conditions.
Over short intervals of time ($\sim 1\,$hour),
the relative detection efficiency as a function of the shower 
direction in local coordinates, ($\theta$,$\phi$),
is largely determined by the array geometry (placement of detectors,
uniformity of terrain, etc.) and is almost constant, and
the time variation of the detector response may be estimated from
the trigger rate.
Therefore,
we separate the detection rate per unit solid
angle in local coordinates, $ N (\theta,\phi,t)$, into
two terms:

\begin{equation} 
N (\theta,\phi,t) \ \ = \ \ D (\theta,\phi) \cdot  R (t) \ \ ,
\label{eq:rate}
\end{equation}

\noindent 
where $ D (\theta,\phi) $  is the efficiency per unit solid angle
of detecting a shower from a given direction in the sky, and
 $R (t) $ is the trigger rate as a function of time.
The factor $ D (\theta,\phi) $ is determined by maps made
from the arrival directions of cosmic ray showers over
given periods of time.
The time dependent term, $R(t)$, is determined from
the arrival times of the actual events.
The expected number of events for a given bin in the sky 
is then determined by integrating $ N (\theta,\phi,t)$
over the time interval in question.
To determine the expected number of events for a bin in
equatorial coordinates, ($\alpha,\delta$), we integrate 
$ N (\theta,\phi,t)$ over the time interval and over local
coordinate space.

More explicitly, the background estimation
is done by the following procedure.
For intervals of $4,200\,$sec,
we accumulate the arrival directions of 
cosmic ray events into 2,700 bins segmented in local coordinate space
(30 bins in $\theta$, 90 bins in $\phi$).
We use the binned data to construct maps of the relative acceptance
of any point in the sky over this time interval.
Separate maps are calculated for each data sample
used in the source search (e.g. all-data and muon-poor data).
To generate simulated background data,
we discard the directional information of an event
and associate the event time with a local coordinate direction
obtained by sampling from the appropriate sky map.
We then
compute artificial values for the
equatorial coordinates 
and determine if this simulated event falls into a search
bin of a source.
By sampling more than once from the sky map for each event time,
we increase the statistics on the 
simulated data sample.
Negligible
systematic bias is introduced by such oversampling.
For this work, we oversample by a factor of ten, an amount that 
is limited only by computational resources.

We have checked that our background estimation method is free from
bias by comparing the detected numbers of events in an
angular bin to our expected number for bins that do not contain
Cygnus X-3 and Hercules X-1.
For each bin, we compute the statistical significance of any
excess or deficit in the number of detected events relative
to the number we predict.
The distribution of these significances is in close agreement
with that expected from statistics, which, because of 
background oversampling,
is dominated by the statistical uncertainty on the number
of detected events.

\subsection{Energy Response}
\label{subsec:energy}
Air shower arrays trigger on the shower size, i.e.
the number of charged particles in the shower at ground level.
For each shower, we determine a shower size
from the particle densities measured in the CASA stations.
For astrophysical interpretations of flux measurements or flux limits,
however, it is necessary to translate from the measured shower
parameters (size and zenith angle) to an estimate of the energy of
the primary particle.
Since there are large fluctuations in shower size 
for showers initiated by particles at fixed energy,
it is difficult for air shower experiments to measure accurately
{\it differential} primary spectra.
Traditionally, therefore, 
flux measurements
have been quoted as {\it integral} intensities above
a fixed energy point.
Although to some degree the energy value at which to quote
the intensity is arbitrary, we desire to use
an energy at which the experiment has a significant
degree of sensitivity.
We chose to quote flux measurements at the {\it median} energy
of the experiment which reduces the dependence of the flux on
the assumed spectral index 
\cite{ref:Ciampa,ref:Gaisser1}.

We estimate the energy response of the experiment by the constant
intensity method, which has been used by other experiments
\cite{ref:Nagano}, as well as by our own group \cite{ref:McKay}.
The constant intensity procedure is described in more detail
elsewhere \cite{ref:Newport}. 
Briefly,
we determine the relationship between shower size and
energy by comparing the detected flux of showers above a given size
to an assumed form of the all-particle cosmic ray spectrum.
The comparison is done on a run-by-run basis to account for
changes in the detector response.
The cosmic ray flux is
derived from
measurements made by other space-borne 
\cite{ref:Asakimori,ref:Swordy}
and ground-based 
\cite{ref:Nagano}
experiments.
The assumed integral cosmic ray intensity above $100\,$TeV is
$6.57 \times 10^{-9}\,$particles cm$^{-2}$ s$^{-1}$ sr$^{-1}$.

We use the relationship between energy and shower size to
determine the most likely energy for each shower coming from the
direction of Cygnus X-3 or Hercules X-1 in a angular bin of fixed
radius.
The medians of the energy distributions determine
the median energies for 
cosmic ray particles from the direction of
Cygnus X-3 and Hercules X-1 that would trigger the experiment and
pass all selection criteria.
By normalizing our energy scale to the cosmic ray flux,
we make the assumption that the primary particle
has the same spectral index as the detected cosmic rays.
This assumption is reasonable when dealing with
sources like Cygnus X-3 and Hercules X-1 in which there are
are no well established measurements of spectral indices.
The median energy of particles from the direction of Cygnus
X-3 is $114\,$TeV, and from Hercules X-1, it is $116\,$TeV.
Since the difference in the energies for the two sources
is negligible, we report our measurements at
a common energy of $115\,$TeV.

\subsection{Search Strategy}
\label{subsec:strategy}
We carry out searches for particle emission from a particular source
by comparing the number of events found
within a circular angular bin
around the source to the number of events estimated by our background
procedure.
The angular bin sizes vary as a function of the number of alerted CASA
stations, as itemized in Table~\ref{tab:bins}.
Source positions (J1992) are taken to be 
$(\alpha,\delta) = (308.04^\circ,40.93^\circ)$ for
Cygnus X-3, and
$(\alpha,\delta) = (254.39^\circ,35.35^\circ)$ for
Hercules X-1.

Separate searches are made based on particle type and energy.
By using the {\it all-data} sample, we are sensitive to any type of
neutral particle that would create air showers.
With the {\it muon-poor} sample, we are specifically sensitive to
the emission of gamma-rays.
We carry out three separate searches with various integral cuts on the
number of alerted CASA stations,
in addition to a search with no cuts.
This procedures takes advantage of the correlation between primary
energy and size (as represented by the number of alerts),
and improves our sensitivity to possible emission
that might be present at either low or high energies.
The data samples selected by cutting on the alert number
and their corresponding
median energies are shown in Table~\ref{tab:energycuts}.

\begin{table}
\begin{center}
\caption{Data samples selected by integral cuts on the number
of CASA alerts.
The cut values are given in the first column and the fractions
of events surviving the cut 
(and within the angular search region)
are shown in the second column.
The median energies for events coming from either Cygnus X-3
or Hercules X-1 are listed in the third column.}
\label{tab:energycuts}
\vspace{10pt}
\begin{tabular}{|crr|}\hline
 Alert Cut &\ \ Event Fraction\ \ &\ \ Median Energy\ \ \\
\hline
None      & 100.00\%\ \ \  &  $115\,$TeV\ \ \ \\
\hline
$\le 10$  &  62.87\%\ \ \  &   $85\,$TeV\ \ \ \\
$> 40$    &   0.58\%\ \ \  & $530\,$TeV\ \ \ \\
$> 80$    &   0.09\%\ \ \  & $1175\,$TeV\ \ \ \\
\hline
\end{tabular}
\end{center}
\end{table}

\section{Results}
\label{sec:results}
We search for evidence of neutral 
(gamma-ray or other) particle emission from Cygnus X-3 and
Hercules X-1.
Separate searches are carried out for steady and
transient emission from either source.
In addition, we search for periodic emission from Cygnus X-3
at the 
$4.8\,$hr
X-ray periodicity and for emission from Cygnus
X-3 that was coincident with the occurrence of large radio flares.
No compelling evidence for emission from either source is found
for all the different searches, and consequently we
set upper limits on the fluxes of particles from the sources.

\subsection{Steady Emission}
\label{subsec:steady}
The numbers of on-source and background events for the various
searches from Cygnus X-3 are shown in Table~\ref{tab:CygEvents}.
The results from similar searches carried out on Hercules X-1
are shown in Table~\ref{tab:HerEvents}.
For each search, we also calculate
the statistical significance of
any excess or deficit in the number of
events observed relative to background by
the prescription of Li and Ma \cite{ref:LiMa}, using
an oversampling factor of 10.
No significant excess is observed for any search from either source.
Therefore, for each search, we calculate
an upper limit, $N_{90}$, on the number of excess events
from the source at the 90\% confidence level
\cite{ref:Helene,ref:PDG}.
Each $N_{90}$ value is converted to a limit on the fractional
excess of events from the source, $f_{90}$, 
by dividing by the estimated number of background events, which is
assumed to represent the background cosmic-ray level.
Since
the $f_{90}$ values are independent of the absolute flux normalization,
they are useful in comparing
results between different experiments.

\begin{table}
\begin{center}
\caption{Steady emission search results for Cygnus X-3 using
the all-data sample (top) and muon-poor sample (bottom).
The number of events observed on-source and the number
expected from background are given in the second the third
columns, respectively.
The fourth column gives the statistical significance of
any excess or deficit.
The 90\% c.l. upper limit on the number of excess events,
$N_{90}$, and the upper limit on the fractional excess,
$f_{90}$, are given in the last two columns.
The methods used to calculate statistical significances
and upper limits are outlined in the text.
The data samples at $85\,$TeV, $530\,$TeV, and $1175\,$TeV
are subsets of the data sample at $115\,$TeV.}
\label{tab:CygEvents}
\vspace{10pt}
\begin{tabular}{|rrrcrc|}
\multicolumn{6}{c}{ All-Data Sample} \\
\hline
Energy & On-Source & Background & Signif. & $N_{90}$ &$f_{90}$ \\
\hline
  $85\,$TeV\ \ & 1119469\ \ & 1119987\ \ & $-0.48\sigma$ & 
        1502.1 &\ \ $1.34\times 10^{-3}$ \\
 $115\,$TeV\ \ & 1780594\ \ & 1781479\ \ & $-0.66\sigma$ & 
        1774.9 &\ \ $9.96\times 10^{-4}$ \\
 $530\,$TeV\ \ &   10286\ \ &   10235\ \ & $+0.49\sigma$ &  
         205.3 &\ \ $2.01\times 10^{-2}$ \\
$1175\,$TeV\ \ &    1583\ \ &    1580\ \ & $+0.08\sigma$ &   
          68.9 &\ \ $4.36\times 10^{-2}$ \\
\hline
\multicolumn{6}{c}{\hphantom{dummy}} \\
\multicolumn{6}{c}{ Muon-Poor Sample} \\
\hline
Energy & On-Source & Background & Signif. & $N_{90}$ &$f_{90}$ \\
\hline
  $85\,$TeV\ \ & 149676\ \ & 149863\ \ & $-0.57\sigma$ & 
        548.1 &\ \ $5.90\times 10^{-4}$ \\
 $115\,$TeV\ \ & 121409\ \ & 121594\ \ & $-0.37\sigma$ & 
        485.4 &\ \ $3.28\times 10^{-4}$ \\
 $530\,$TeV\ \ &   20\ \ &   21.0\ \ & $-0.21\sigma$ &  
         8.2 &\ \ $9.47\times 10^{-4}$ \\
$1175\,$TeV\ \ &    1\ \ &    0.6\ \ & $+0.44\sigma$ &   
          3.5 &\ \ $2.67\times 10^{-3}$ \\
\hline
\end{tabular}
\end{center}
\end{table}

\begin{table}
\begin{center}
\caption{Steady emission search results for Hercules X-1 using
the all-data sample (top) and muon-poor sample (bottom).}
\label{tab:HerEvents}
\vspace{10pt}
\begin{tabular}{|rrrcrc|}
\multicolumn{6}{c}{ All-Data Sample} \\
\hline
Energy & On-Source & Background & Signif. & $N_{90}$ &$f_{90}$ \\
\hline
  $85\,$TeV\ \ & 1058904\ \ & 1057583\ \ & $+1.12\sigma$ & 
        2738.1 &\ \ $2.59\times 10^{-3}$ \\
 $115\,$TeV\ \ & 1681708\ \ & 1681392\ \ & $+0.23\sigma$ & 
        2387.6 &\ \ $1.42\times 10^{-3}$ \\
 $530\,$TeV\ \ &   9579\ \ &     9532\ \ & $+0.46\sigma$ &  
         196.5 &\ \ $2.06\times 10^{-2}$ \\
$1175\,$TeV\ \ &    1419\ \ &    1459\ \ & $-0.98\sigma$ &   
          44.0 &\ \ $3.02\times 10^{-2}$ \\
\hline
\multicolumn{6}{c}{\hphantom{dummy}} \\
\multicolumn{6}{c}{ Muon-Poor Sample} \\
\hline
Energy & On-Source & Background & Signif. & $N_{90}$ &$f_{90}$ \\
\hline
  $85\,$TeV\ \ & 139580\ \ & 139670\ \ & $-0.24\sigma$ & 
        577.1 &\ \ $6.57\times 10^{-4}$ \\
 $115\,$TeV\ \ & 113360\ \ & 113244\ \ & $+0.37\sigma$ & 
        643.4 &\ \ $4.62\times 10^{-4}$ \\
 $530\,$TeV\ \ &   14\ \ &   16.8\ \ & $-0.67\sigma$ &  
         6.3 &\ \ $7.96\times 10^{-4}$ \\
$1175\,$TeV\ \ &    0\ \ &    0.5\ \ & $-0.98\sigma$ &   
          2.3 &\ \ $1.90\times 10^{-3}$ \\
\hline
\end{tabular}
\end{center}
\end{table}

\begin{figure} 
\centerline{\ \psfig{figure=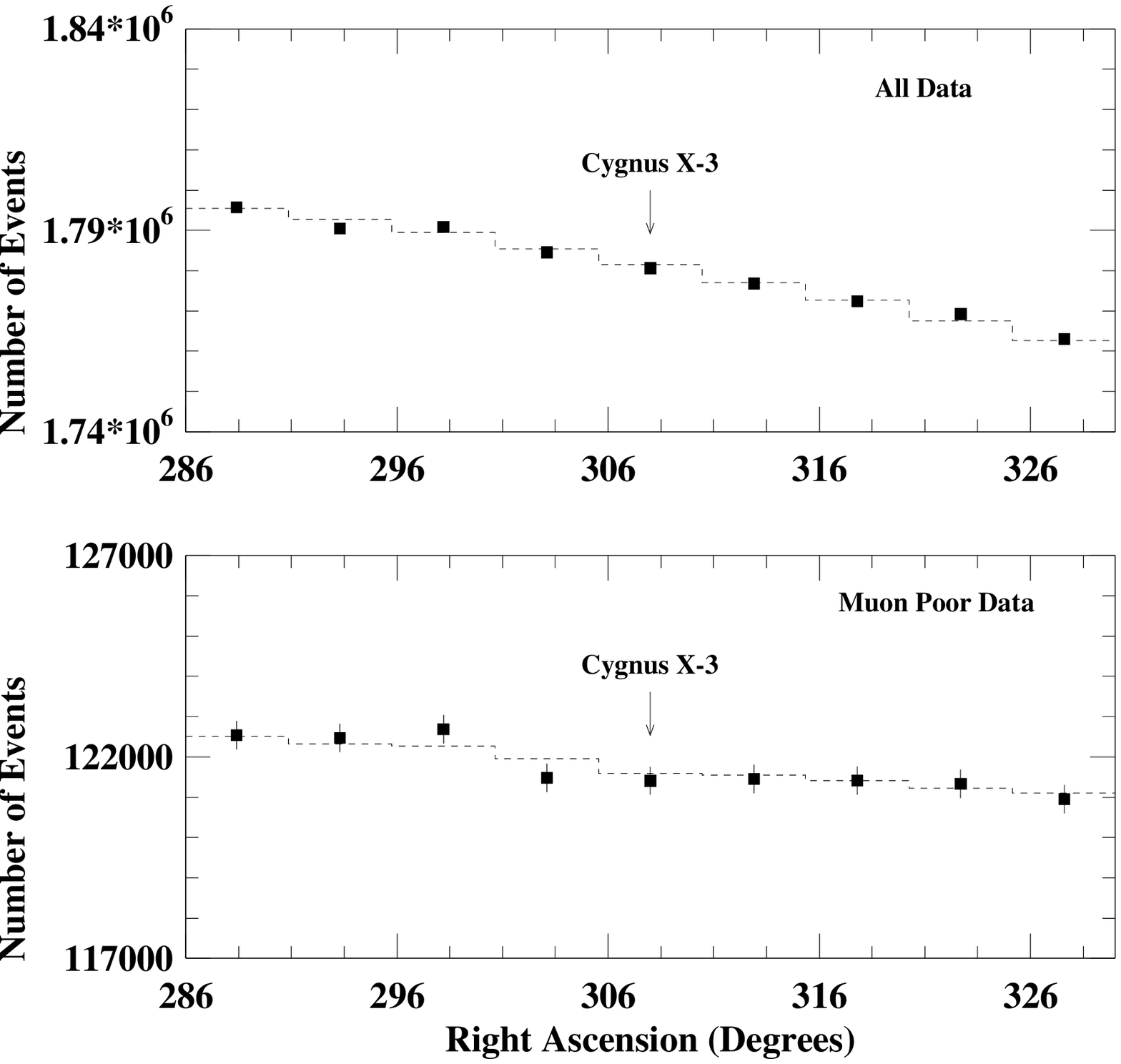,height=17.0cm}}
\caption{
Scans in right ascension for a band of
constant declination centered
on Cygnus X-3, at a median energy of 115 TeV.
The data points correspond to the numbers of events observed
in each bin; the dashed histogram is the expected background level.
Top plot shows the all-data sample; bottom plot shows the
muon-poor sample.
Note that the scale on the vertical axis has been highly 
expanded and zero-suppressed.}
\label{fig:CygScan}
\end{figure}

\begin{figure} 
\centerline{\ \psfig{figure=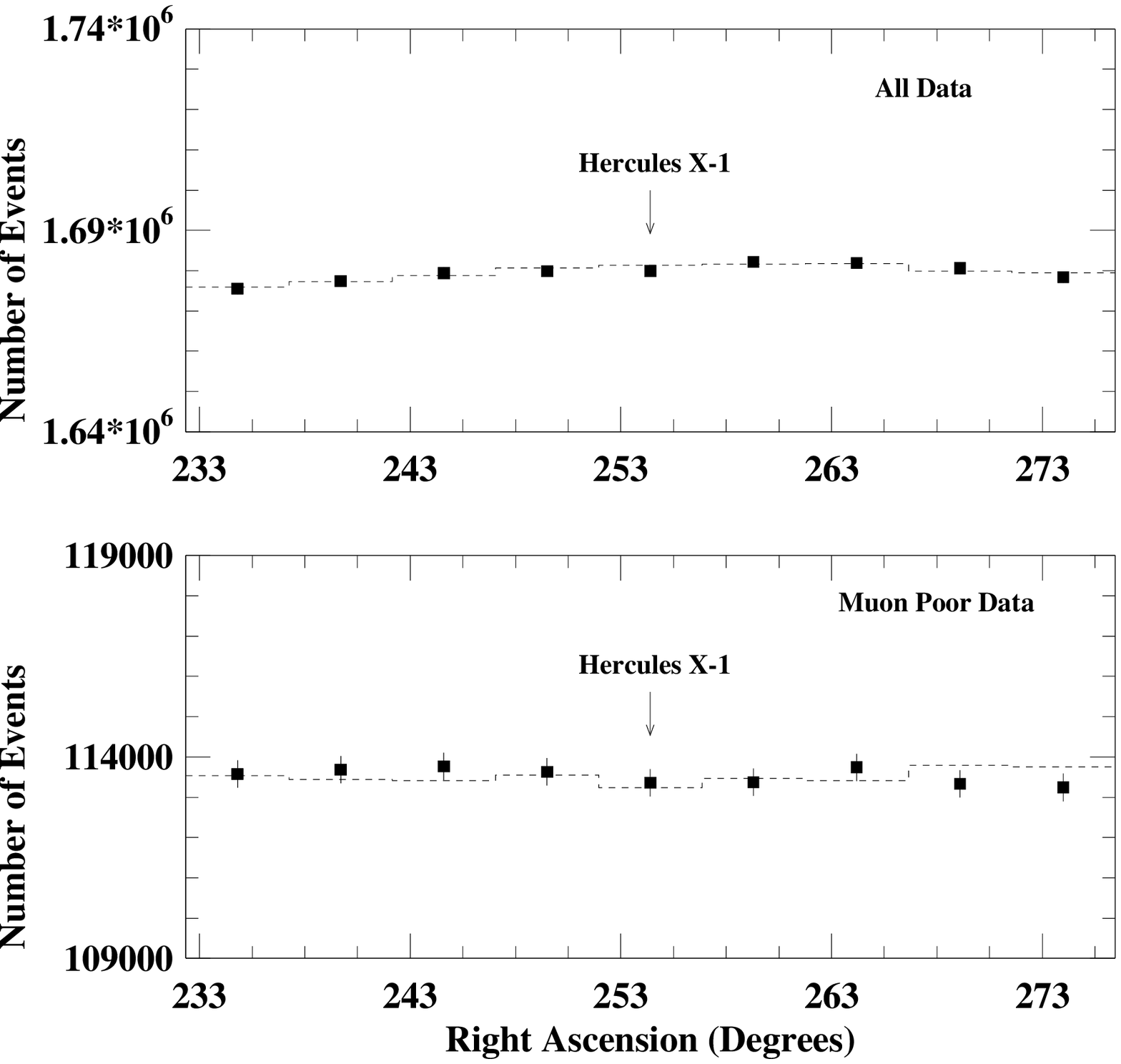,height=17.0cm}}
\caption{
Scans in right ascension for a band of
constant declination centered
on Hercules X-1, at a median energy of 115 TeV.
The data points correspond to the numbers of events observed
in each bin; the dashed histogram is the expected background level.
Top plot shows the all-data sample; bottom plot shows the
muon-poor sample.
Note that the scale on the vertical axis has been highly 
expanded and zero-suppressed.}
\label{fig:HerScan}
\end{figure}

Figure~\ref{fig:CygScan} shows scans in right ascension for
bands of declination centered on Cygnus X-3 for the all-data
and muon-poor samples.
No significant excess above background 
is seen in either sample for the bin
containing Cygnus X-3.
The background estimation 
agrees well with the data in the off-source region.
Similar scans for Hercules X-1 are shown in 
Figure~\ref{fig:HerScan}, and
again the background estimation agrees well with the observed data
and no excesses are seen.

\subsubsection{Flux Limit Calculation}
\label{subsec:flux}
In the absence of a statistically significant excess
from either Cygnus X-3 or Hercules X-1, we set upper limits
on the flux of particles from each source.
Separate limits are set for neutral and gamma-ray primaries.
For gamma-ray primaries, the 90\% c.l. upper limit, 
$\Phi_\gamma (E)$,
on the integral flux is calculated from the measured
fractional excess by normalizing to the cosmic ray flux:

\begin{equation}
\Phi_\gamma\ (E) \ \ = \ \  
   {  { f_{90}\ \bar{\Omega} } \over {\epsilon\ R_\gamma} }
    \ J (E)\ .
\label{eq:limit}
\end{equation}

\noindent Here, 
$\bar{\Omega}$ is the mean solid angle used in the search,
$\epsilon$ is the fraction of events that would pass cuts
and fall into the search bin,
$J (E)$ is the integral cosmic ray intensity above energy $E$, 
and $R_\gamma$ is a factor which accounts for the relative trigger
efficiency for gamma-rays as opposed to cosmic rays.
The value of $\bar{\Omega}$ ranges from $5.74\times 10^{-3}\,$sr for
the lowest energy data set to
$1.60\times 10^{-4}\,$sr for the highest energy data set.
The $\epsilon$ factor accounts for the fraction of gamma-rays that
would end up in the angular search bin (0.72) and the fraction
that would pass the muon-poor selection
criterion (Table~\ref{tab:rmucut}).
The value of $R_\gamma$ was determined by Monte Carlo
simulations to be 1.6.

To determine an upper limit on the integral
flux of any neutral particle from
a source, $\Phi_N (E)$, we use
Eq.~\ref{eq:limit}, except $\epsilon$ is now 0.72
and $R_\gamma$ is 1.0.
In this calculation, we assume that the neutral particle 
would interact in the atmosphere to create air showers
in a similar manner to cosmic rays.

Table~\ref{tab:CygHerLimits} gives the flux limits obtained from the
various searches for steady emission from 
the two sources.
Limits are not calculated for the data samples with median energies
of $85\,$TeV because these are {\it not} integral energy samples.

\begin{table}
\begin{center}
\caption{Flux limits from searches for steady emission from 
Cygnus X-3 (top) and
Hercules X-1 (bottom). The second and third columns give the 90\% c.l. upper
limit on the integral flux of any neutral or gamma-ray particles
from the source, respectively.
The units of flux are particles cm$^{-2}$ sec$^{-1}$.}
\label{tab:CygHerLimits}
\vspace{10pt}
\begin{tabular}{|rcc|}
\multicolumn{3}{c}{ Cygnus X-3} \\
\hline
Energy & $\Phi_N\ (E) $ & $\Phi_\gamma\ (E)$ \\
\hline
 $115\,$TeV\ \ & $2.20 \times 10^{-14}$ & $6.26 \times 10^{-15}$ \\
 $530\,$TeV\ \ & $1.43 \times 10^{-15}$ & $1.21 \times 10^{-16}$ \\
$1175\,$TeV\ \ & $1.04 \times 10^{-15}$ & $5.19 \times 10^{-17}$ \\
\hline
\multicolumn{3}{c}{\hphantom{dummy}} \\
\multicolumn{3}{c}{ Hercules X-1} \\
\hline
Energy & $\Phi_N\ (E) $ & $\Phi_\gamma\ (E)$ \\
\hline
 $115\,$TeV\ \ & $3.04 \times 10^{-14}$ & $8.55 \times 10^{-15}$ \\
 $530\,$TeV\ \ & $2.87 \times 10^{-15}$ & $9.75 \times 10^{-17}$ \\
$1175\,$TeV\ \ & $6.91 \times 10^{-16}$ & $3.56 \times 10^{-17}$ \\
\hline
\end{tabular}
\end{center}
\end{table}

\subsection{Transient Emission}
\label{subsec:transient}
We search for transient emission of particles
from Cygnus X-3 and Hercules X-1 on daily (single transit)
time scales.
For each transit of the source, we compare the number of 
on-source events to the number of expected background events
and calculate a significance based on
the prescription of Li and Ma \cite{ref:LiMa}.
We require the live time fraction during the transit to be
at least 0.20 to remove transits in which the experiment was
operational for only a small fraction of the time.
For each source,
we make separate studies of the transit significances for the
all-data and muon-poor samples, corresponding to possible emission
from any neutral and gamma-ray particles, respectively.

\begin{figure} 
\centerline{\ \psfig{figure=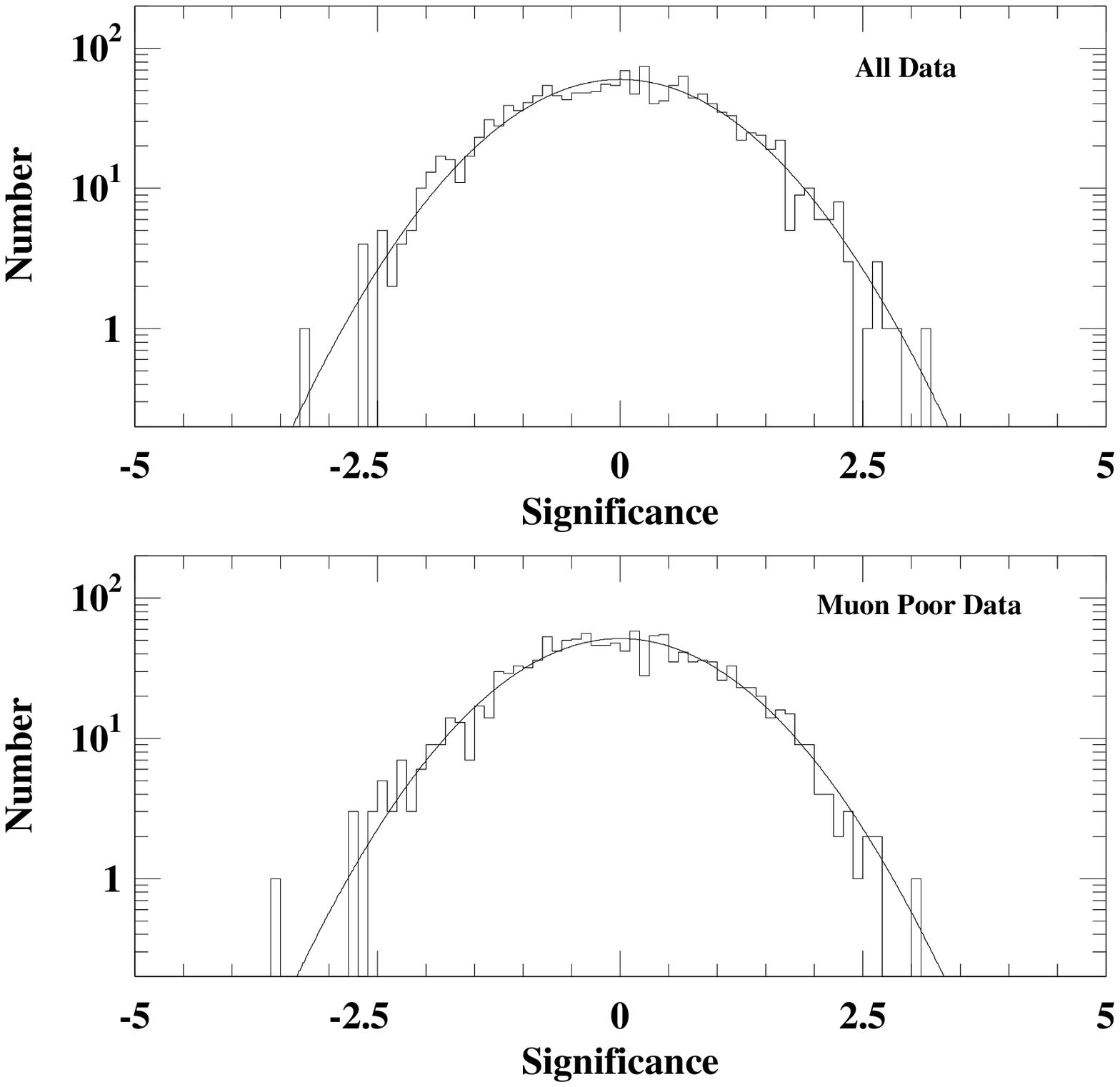,height=17.0cm}}
\caption{
Distribution of transit significances for 
Cygnus X-3 for the all-data (top) and muon-poor (bottom) samples.
The curves are unit-width Gaussian distributions with zero mean.
For comparison, the fitted mean (standard deviation) of the
all-data sample is $-0.040 \pm 0.031$ ($1.011 \pm 0.021$) and
the fitted mean (standard deviation) of the muon-poor sample
is $-0.011 \pm 0.032$ ($0.992 \pm 0.022$).}
\label{fig:Cyg_Trans}
\end{figure}

For Cygnus X-3, the number of good transits in the
all-data sample is 1500.
In the muon-poor sample, it is 1291.
The distributions of significances for 
the two samples of Cygnus X-3 transits
are shown in Figure~\ref{fig:Cyg_Trans}.
Each
distribution agrees well with a Gaussian
distribution
of mean zero and unit width.
There is no
evidence for any excess of events at high values of significance
(either positive or negative).

\begin{figure} 
\centerline{\ \psfig{figure=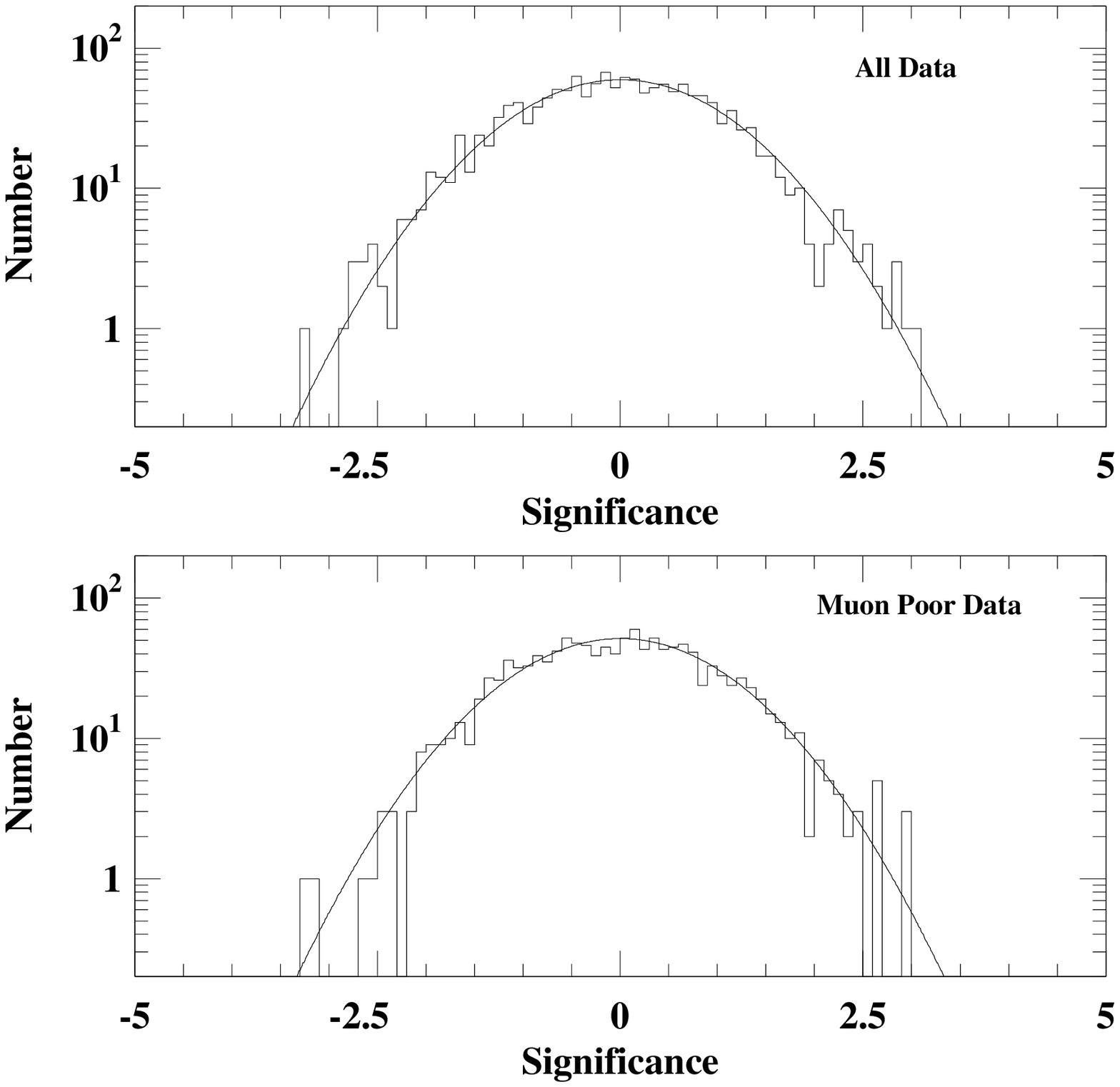,height=17.0cm}}
\caption{
Distribution of transit significances for 
Hercules X-3 for the all-data (top) and muon-poor (bottom) samples.
The curves are unit-width Gaussian distributions with zero mean.
For comparison, the fitted mean (standard deviation) of the
all-data sample is $-0.032 \pm 0.031$ ($0.973 \pm 0.020$) and
the fitted mean (standard deviation) of the muon-poor sample
is $0.030 \pm 0.033$ ($0.976 \pm 0.022$).}
\label{fig:Her_Trans}
\end{figure}

For Hercules X-1, there are 1492 good transits in the all-data
sample, and 1271 good transits in the muon-poor sample.
The significance distributions for Hercules X-1 are shown in
Figure~\ref{fig:Her_Trans}, and again, no evidence for 
significant excesses exists.

Based on the lack of statistically significant excesses,
we place limits on the daily
fluxes of neutral and gamma-ray particles
from Cygnus X-3 and Hercules X-1.
These limits are calculated by a similar procedure as used
for the steady searches.
The limit values depend on the actual statistical significance
of the search on a given day, and also on the 
epoch of data taking.
As shown in Table~\ref{tab:array}, the size of the experiment has
changed with time, and the sensitivity changed accordingly.
In Table~\ref{tab:daily}, we give typical daily flux
limits for the two sources for different epochs of the experiment.
Since the numbers of events detected per transit are the
same for the two sources to within 5\%, the limits for
Cygnus X-3 and Hercules X-1 are virtually identical.
Typical 90\% c.l. limits on the integral flux
using the full experiment
are $\Phi_N ( E > 115\,{\rm TeV} ) < 9.7\times 10^{-13}$
neutral particles cm$^{-2}$ sec$^{-1}$ and
 $\Phi_\gamma ( E > 115\,{\rm TeV} ) < 2.0\times 10^{-13}$
photons cm$^{-2}$ sec$^{-1}$.

\begin{table}
\begin{center}
\caption{Typical daily
upper flux limits (90\% c.l.)
for emission of neutral and
gamma-ray particles from Cygnus X-3 and Hercules X-1.
The flux limits are calculated for two different epochs assuming
the same number of on-source events as off-source.
The third column gives
the typical number of
events observed on-source during the different epochs.
Epoch I corresponds to March 1990 to October 1990.
Epoch II corresponds to January 1992 to August 1993.
Flux limits for the remaining periods of time of operation
are close to those for Epoch II.
Units of flux are particles cm$^{-2}$ sec$^{-1}$.}
\label{tab:daily}
\vspace{10pt}
\begin{tabular}{|rcrrc|}
\multicolumn{5}{c}{Epoch I} \\
\hline
Energy & Particle &\ Events\ \ & $f_{90}\ \ $ & $\Phi_{daily}(E)$ \\
\hline
$115\,$TeV  & Any      & 575    & 0.071  &  $\ \ 1.6\times 10^{-12}$ \\
$530\,$TeV  & Any      &  3.2   & 1.34   &  $\ \ 1.9\times 10^{-13}$ \\
$1175\,$TeV & Any      &  0.56  & 4.11   &  $\ \ 9.6\times 10^{-14}$ \\
$115\,$TeV  & $\gamma$-ray &  45    & 0.026  &  $\ \ 3.6\times 10^{-13}$ \\
\hline 
\multicolumn{5}{c}{\hphantom{dummy}} \\
\multicolumn{5}{c}{Epoch II} \\
\hline
Energy & Particle &\ Events\ \ & $f_{90}\ \ $ & $\Phi_{daily}(E)$ \\
\hline
$115\,$TeV  & Any     & 1450    & 0.044  & $\ \ 9.7\times 10^{-13}$ \\
$530\,$TeV  & Any     &  8.1   & 0.76   &  $\ \ 1.1\times 10^{-13}$ \\
$1175\,$TeV & Any     &  1.3  &  2.56   &  $\ \ 5.9\times 10^{-14}$ \\
$115\,$TeV  & $\gamma$-ray &  96    & 0.015  &  $\ \ 2.0\times 10^{-13}$ \\
\hline
\end{tabular}
\end{center}
\end{table}

We have also carried out searches for transient emission 
on the shorter time scale of 
$0.5\,$hr.
Here, we compare the number of events observed on-source
to the expected background level for ten 
$0.5\,$hr
time intervals on either side of the time of source culmination.
The typical number of on-source events, and therefore the flux
sensitivity, depends strongly on the source zenith angle.
For example, for an overhead source
near culmination, the experiment observes
$\sim 175$ events per 
$0.5\,$hr,
whereas at four hours from
culmination the rate is $\sim 15$ events per 
$0.5\,$hr.
Regardless of the rate, for each 
$0.5\,$hr
interval, we calculate the significance in
the number of on-source events relative to the background
and combine all such significances into
a single distribution.
The resulting
significance distributions are consistent with those expected
from background processes for both sources in both the all-data
and muon-poor samples.
The typical 90\% c.l. upper limits on the fluxes
from either source are
$\Phi_N ( E > 115\,{\rm TeV} ) < 3.1\times 
10^{-12}$ neutral particles cm$^{-2}$ sec$^{-1}$
and
$\Phi_\gamma ( E > 115\,{\rm TeV} ) < 7.1\times 10^{-13}$
photons cm$^{-2}$ sec$^{-1}$
for $0.5\,$hr
periods within one hour of culmination.

\subsubsection{Cygnus X-3 Radio Flares}
\label{subsec:radio}
We study showers from the direction of Cygnus X-3 during
the occurrence of large radio flares at the source.
We define large flares
as those times when the radio
output at 8.3 GHz exceeded 2 Jy, a level which is
two orders of magnitude above the typical quiescent level.
During the period of CASA-MIA operations, there
were six large flares, as listed in Table~\ref{tab:flares}
\cite{ref:Waltman1,ref:Waltman2}.

\begin{table}
\begin{center}
\caption{Large radio flares of Cygnus X-3 from 1990 to 1995,
coincident with the operational time of CASA-MIA.
The flare number is an arbitrary index used for this work.
The peak radio flux values (8.3 GHz)
come from 
\protect\cite{ref:Waltman1,ref:Waltman2}.
The March 1994  flare was actually a prolonged event that
extended for the ten days following March 1, 1994.}
\label{tab:flares}
\vspace{10pt}
\begin{tabular}{|crr|}\hline
Flare & Date\ \ \ \ \ \ \ \ & Peak Flux (Jy) \\
\hline
1 &\ \ Aug. 15, 1990\ \ &  7.5\ \ \ \ \ \ \ \ \ \\
2 & Oct. 05, 1990  & 10.2\ \ \ \ \ \ \ \ \ \\
3 & Jan. 21, 1991  & 14.8\ \ \ \ \ \ \ \ \ \\
4 & Sep. 04, 1992  &  4.1\ \ \ \ \ \ \ \ \ \\
5 & Feb. 20, 1994  &  4.9\ \ \ \ \ \ \ \ \ \\
6 & Mar. 09, 1994  &  5.2\ \ \ \ \ \ \ \ \ \\
\hline
\end{tabular}
\end{center}
\end{table}

We examine the daily significances for Cygnus X-3 on
the day of each large flare, as well as on the
day preceeding and following each flare.
Table~\ref{tab:flare_results} lists the numbers of
observed events, the expected background, and the
Li-Ma significances
for the examined days.
There is no
compelling evidence for any statistical excess in the 
observed number of
events from Cygnus X-3 for either the all-data or muon-poor samples.
On one day (Feb. 20, 1994) the Li-Ma
significance is 2.26$\sigma$ for the all-data sample.
The probability that we would get a day with this level of
significance or greater is 28.2\% after accounting for the
fifteen days in which we searched.
In addition, on this same day, there is no evidence for any
excess in the muon-poor data, while
we would expect the statistical
significance to increase by a factor of 3.3 if it were
due to a gamma-ray signal.
Table~\ref{tab:flare_results} also lists the derived 
fractional excess values, $f_{90}$, as well as the upper limits
to the integral flux of neutral or gamma-ray particles from
Cygnus X-3 during the flares.

\begin{table}
\begin{center}
\caption{CASA-MIA search results for emission from
Cygnus X-3 near the time of large radio flares.
The flare numbers are defined in 
Table~\protect\ref{tab:flares}.
The $-$ and $+$ designations refer to the days preceeding
and following the flare day, respectively.
The significances (columns 4 and 8) are 
standard deviation values calculated using
the prescription of Li and Ma 
\protect\cite{ref:LiMa}.
The last two columns give the 90\% c.l. upper limits on the
integral flux above $115\,$TeV
of any neutral particle and gamma-rays, respectively,
in units of 
$10^{-12}$ particles cm$^{-2}$ sec$^{-1}$.
Entries having only a dash indicate the absence of
any usable data.}
\label{tab:flare_results}
\vspace{10pt}
\begin{tabular}{|c|rrrr|rrrr|cc|}\hline
\ \ & \multicolumn{4}{c|}{ All-Data Sample} &
\multicolumn{4}{c|}{ Muon-Poor Sample} &
\multicolumn{2}{c|}{ Flux Limits}  \\
\hline
Flare & On & Back & Signif. & $f_{90}$(\%) &
        On & Back & Signif. & $f_{90}$(\%) &
        $\Phi_N(E)$ & $\Phi_\gamma(E) $ \\
\hline
$-$& 226 & 241.2 & $-0.94$ &  7.8\ \ & --\ \ & --\ \ & --\ \ & --\ \ & 1.7 & 
  --\ \ \\
 1 & 249 & 249.5 & $-0.03$ & 10.8\ \ & --\ \ & --\ \ & --\ \ & --\ \ & 2.4 & 
  --\ \ \\
$+$& 252 & 248.1 & $+0.25$ & 12.1\ \ & 22 & 19.4 & $+0.58$ & 5.2\ \ & 
2.7 & 0.73 \\
\hline
$-$& 850 & 863.1 & $-0.43$ &  4.9\ \ & 121 & 113.4 & $+0.67$ & 3.4\ \ & 
1.1 & 0.48 \\
 2 & 884 & 849.1 & $+1.13$ &  9.0\ \ &  94 & 111.7 & $-1.65$ & 1.5\ \ & 
2.0 & 0.21 \\
$+$& 872 & 901.4 & $-0.94$ &  3.9\ \ & 103 & 124.5 & $-1.90$ & 1.4\ \ & 
0.9 & 0.20 \\
\hline
$-$&1279 &1260.1 & $+0.51$ &  5.8\ \ & 227 & 230.7 & $-0.23$ & 2.3\ \ & 
1.3 & 0.32 \\
 3 &1297 &1267.3 & $+0.80$ &  6.4\ \ & 255 & 235.5 & $+1.19$ & 4.0\ \ & 
1.4 & 0.56 \\
$+$&1265 &1235.2 & $+0.81$ &  6.6\ \ & 216 & 227.2 & $-0.71$ & 1.9\ \ & 
1.5 & 0.27 \\
\hline
$-$& 884 & 992.8 & $+0.04$ &  5.8\ \ &  59 &  50.2 & $+1.15$ & 2.8\ \ & 
1.3 & 0.39 \\
 4 & 711 & 756.9 & $-1.61$ &  3.4\ \ &  45 &  53.4 & $-1.13$ & 1.4\ \ & 
0.8 & 0.20 \\
$+$& 735 & 773.9 & $-1.35$ &  3.7\ \ &  30 &  43.4 & $-2.07$ & 0.9\ \ & 
0.8 & 0.13 \\
\hline
$-$&1708 &1734.0 & $-0.60$ &  3.2\ \ & 119 & 126.9 & $-0.68$ & 1.1\ \ & 
0.7 & 0.15 \\
 5 &1692 &1596.2 & $+2.26$ &  9.4\ \ & 119 & 124.4 & $-0.47$ & 1.2\ \ & 
2.1 & 0.17 \\
$+$&1485 &1447.1 & $+0.95$ &  6.4\ \ & 127 & 119.1 & $+0.68$ & 2.1\ \ & 
1.4 & 0.29 \\
\hline
$-$&1382 &1375.2 & $+0.18$ &  4.9\ \ &  92 & 102.6 & $-1.02$ & 1.1\ \ & 
1.1 & 0.15 \\
 6 &1212 &1225.3 & $-0.36$ &  4.2\ \ &  90 &  95.1 & $-0.50$ & 1.4\ \ & 
0.9 & 0.20 \\
$+$&1390 &1391.7 & $-0.06$ &  4.4\ \ & 108 &  99.1 & $+0.84$ & 2.1\ \ & 
1.0 & 0.29 \\
\hline
\end{tabular}
\end{center}
\end{table}

\subsection{Periodic Emission}
\label{subsec:periodic}
Several previous observations of Cygnus X-3 claimed evidence for
steady emission correlated with the 
$4.8\,$hr
X-ray periodicity of the source.
For this reason, we carry out a search for such emission using the
entire CASA-MIA data set.
The event arrival times (UT) are corrected to the barycenter
of the solar system using the JPL DE200 planetary ephemeris
\cite{ref:Standish}.
The corrected times are folded with the 
$4.8\,$hr
X-ray ephemeris
of van der Klis and Bonnet-Bidaud \cite{ref:vanderKlis2}.
A slight correction is made for newer X-ray data from the ASCA
satellite, as reported by Kitamoto {\it et al.} \cite{ref:Kitamoto}.
Each event is then assigned a phase value
in the interval (0,1) representing the fraction of a period 
that the event is from the X-ray minimum. The phase values
are accumulated in twenty bins of 0.05 phase units each for
both the on-source and
generated background events.

\begin{figure} 
\centerline{\ \psfig{figure=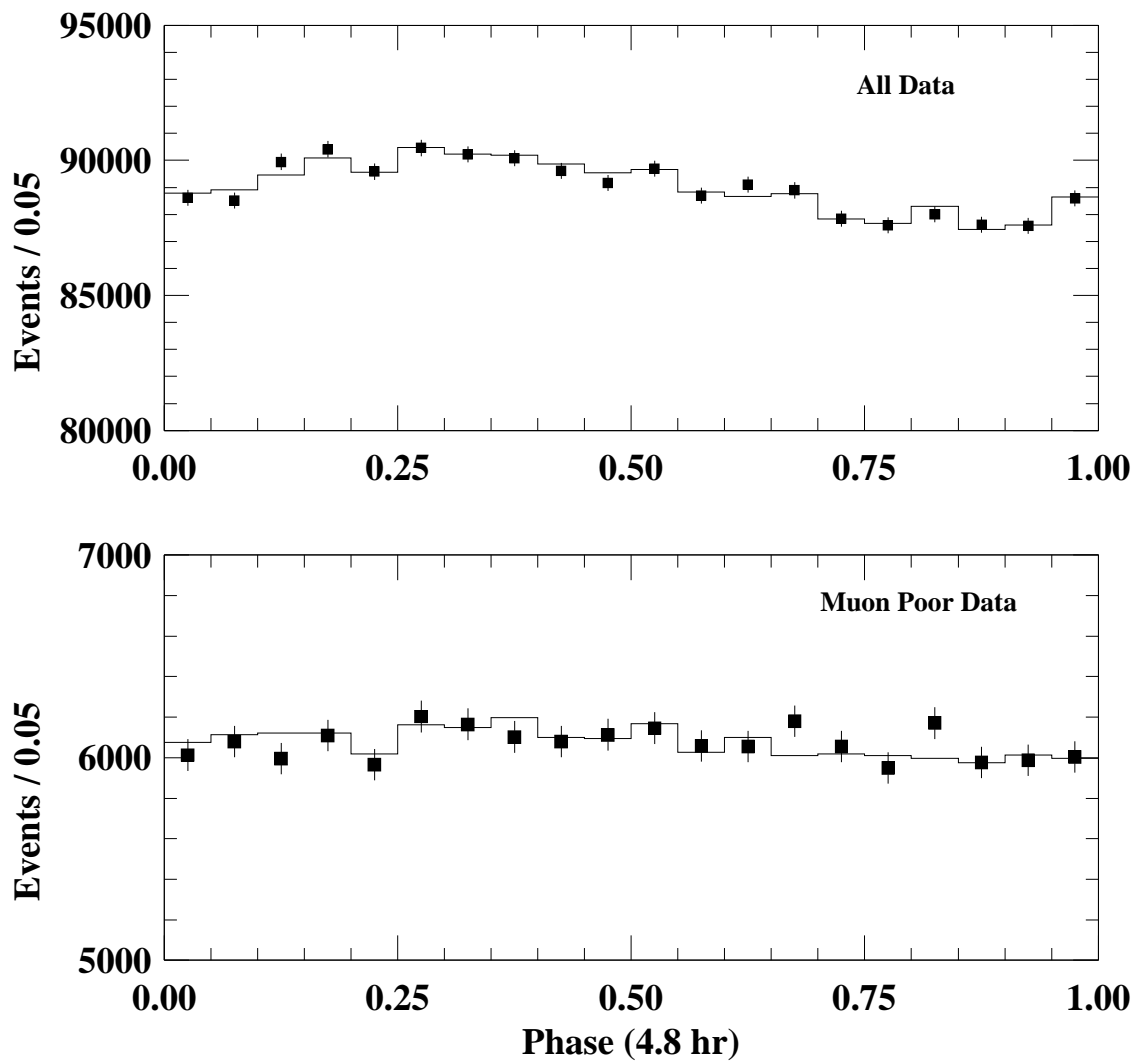,height=17.0cm}}
\caption{
Phase distributions for events from Cygnus X-3 based
on the 
$4.8\,$hr
X-ray periodicity.
Data points show the numbers of on-source events in
each phase bin for the all-data (top) and muon-poor (bottom)
samples.
Histograms show the expected background levels.}
\label{fig:CygPhase}
\end{figure}

Figure~\ref{fig:CygPhase} shows the 
$4.8\,$hr
periodicity
distribution of events from
the direction of Cygnus X-3 for the all-data and muon-poor samples.
Also shown is the phase distribution expected from the background events.
No compelling excesses are seen at any particular phase interval
for either sample.
We carry out similar periodicity analyses using data
at higher energies selected by the number of alerted CASA stations.
These searches also do not indicate any significant excesses
at any phase interval.
In Table~\ref{tab:CygPhase}, we list flux limits for the various
searches at the phase intervals (0.2,0.3) and (0.6,0.7).
These intervals were ones in which
numerous earlier experiments had reported detections.

\begin{table}
\begin{center}
\caption{CASA-MIA search results for 
$4.8\,$hr
periodic emission from
Cygnus X-3.
Flux limits are given for selected phase intervals in which
earlier experiments had reported detections.
Columns 3 and 4 give the 90\% c.l. upper limits to the integral
flux of neutral and gamma-ray particles, respectively, in
units of particles cm$^{-2}$ sec$^{-1}$.
A blank entry corresponds to a data set having
insufficient data with which to calculate a limit.}
\label{tab:CygPhase}
\vspace{10pt}
\begin{tabular}{|crcc|}\hline
Phase Interval & Energy  & $\Phi_N(E)$ & $\Phi_\gamma(E)$ \\
\hline
$0.2 - 0.3 $   & $115\,$TeV\ &  $8.9\times 10^{-14}$ & $2.3\times 10^{-14}$ \\
               & $530\,$TeV &  $4.5\times 10^{-15}$ & $6.9\times 10^{-16}$ \\
               &$1175\,$TeV &  $3.5\times 10^{-15}$ & --\ \ \\
\hline
$0.6 - 0.7 $   & $115\,$TeV &  $1.4\times 10^{-13}$ & $3.5\times 10^{-14}$ \\
               & $530\,$TeV &  $3.8\times 10^{-15}$ & $3.4\times 10^{-16}$ \\
               &$1175\,$TeV &  $6.5\times 10^{-15}$ & --\ \ \\
\hline
\end{tabular}
\end{center}
\end{table}

\section{Comparison with Other Results}
\label{sec:comparison}
As described earlier, the many
detections of Cygnus X-3 and Hercules X-1 by experiments
operating between 1975 and 1990 varied greatly in their
characteristics.
Some results were steady and some were episodic;
some exhibited apparent periodicity and others did not.
We thus choose to compare the results of this work to a
generalized picture of the earlier results and to more recent
work.

\subsection{Cygnus X-3}
\label{subsec:cygnus_comparison}
\begin{figure} 
\centerline{\ \psfig{figure=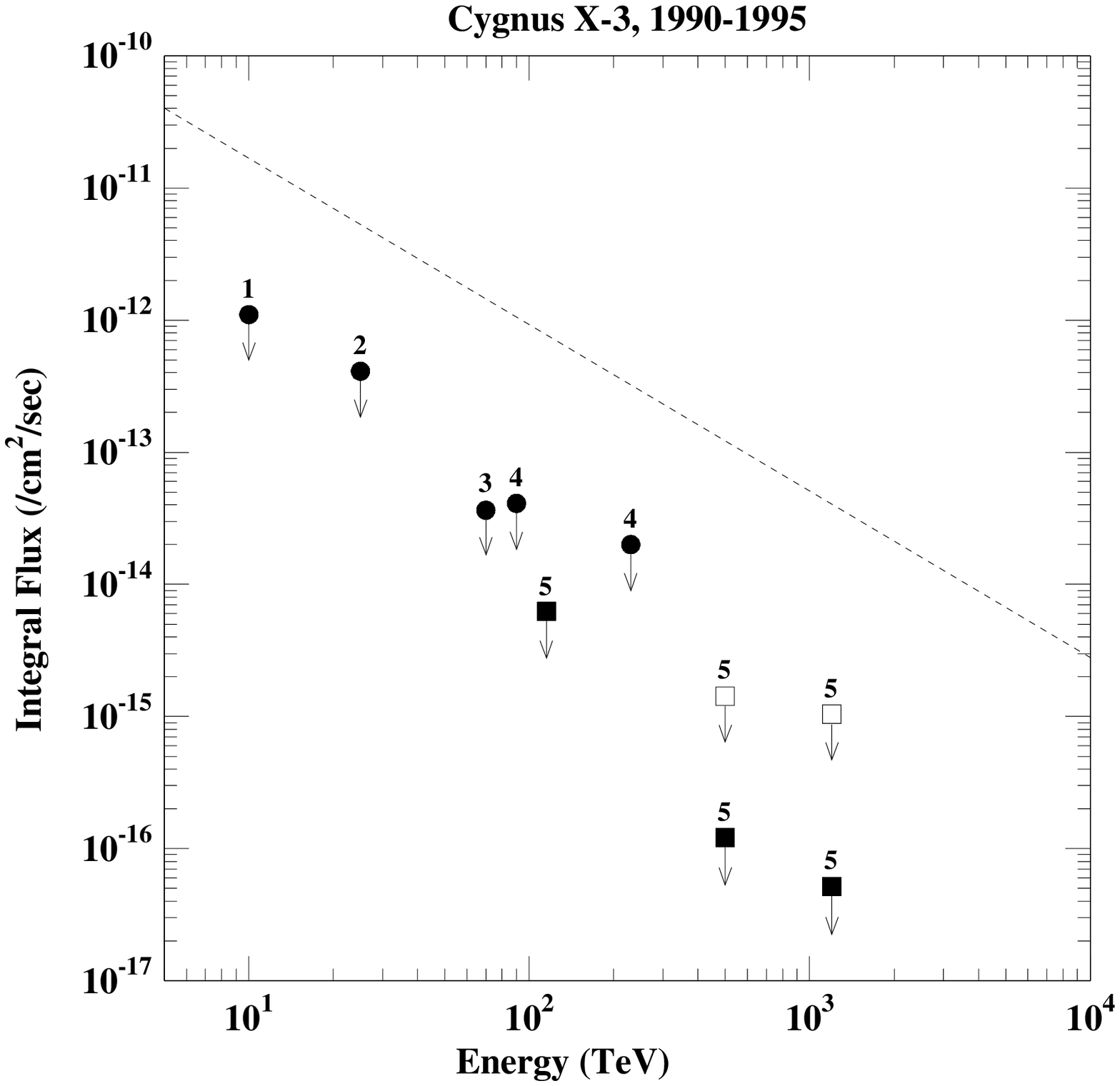,height=17.0cm}}
\caption{
Flux limits reported between 1990 and 1995 on the
steady emission of particles from Cygnus X-3.
The squares (5) represent the results of this work.
Open squares indicate limits on the emission of any neutral
particle that creates air showers.
Filled
squares indicate limits on the emission of gamma-rays.
The circles (1-4) represent results from other experiments:
1. Tibet \protect\cite{ref:Amenomori},       
2. HEGRA \protect\cite{ref:Karle}, 
3. CYGNUS \protect\cite{ref:Alexandreas4},
and 4. EAS-TOP \protect\cite{ref:Aglietta}.
The dashed curve is the approximate power law fit
to early results (reproduced from Figure~\protect\ref{fig:OldCyg}).}
\label{fig:NewCyg}
\end{figure}

\begin{figure} 
\centerline{\ \psfig{figure=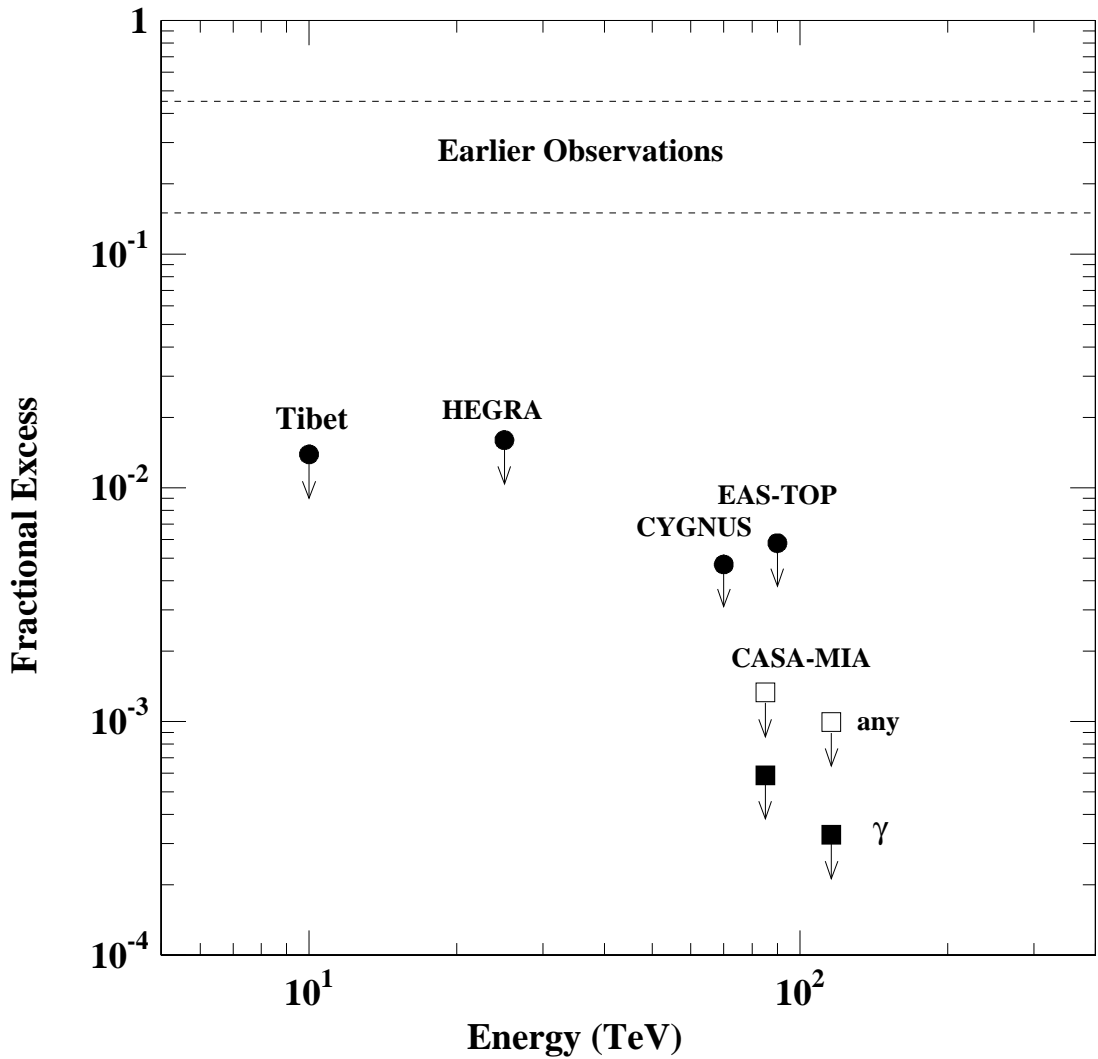,height=17.0cm}}
\caption{
Limits reported between 1990 and 1995 on the
fractional excess of gamma-rays from Cygnus X-3
relative to the cosmic ray background.
The squares represent the results of this work.
Open squares indicate limits on the emission of any neutral
particle that creates air showers.
Filled
squares indicate limits on the emission of gamma-rays.
The circles represent results from other experiments:
Tibet \protect\cite{ref:Amenomori},  
HEGRA \protect\cite{ref:Karle}, 
CYGNUS \protect\cite{ref:Alexandreas4},
and EAS-TOP \protect\cite{ref:Aglietta}.
The dashed lines indicate the range 
of fractional excess values corresponding to the fluxes
reported by earlier experiments.}
\label{fig:CygFract}
\end{figure}

In Figure~\ref{fig:NewCyg}, we 
plot the flux limits reported here on 
steady emission from Cygnus X-3.
We also show
published results from other experiments
using data taken at times which overlap our observation period.
The other results come from the 
Tibet air shower array in Yangbajing, China 
\cite{ref:Amenomori}, the
CYGNUS array in New Mexico, USA
\cite{ref:Alexandreas4},
the EAS-TOP array at Gran Sasso, Italy
\cite{ref:Aglietta},
and the HEGRA experiment on the Canary Island La Palma
\cite{ref:Karle}.    
We do not show earlier results from data taken by 
a portion of our experiment
in 1989 \cite{ref:Cronin} or the results of our all-sky survey
for northern hemisphere point sources using data taken
in 1990-1991 \cite{ref:McKay}.
In Figure~\ref{fig:CygFract}, we show a similar comparison of
the limits on the fractional excess of events from Cygnus X-3 relative to
the cosmic ray background.

The data from the recent experiments are consistent; no steady
emission of ultra-high energy
particles (gamma-ray or otherwise) has been
detected from Cygnus X-3 at levels which are considerably lower
than earlier reports.
At TeV energies, the results from the Whipple Telescope
\cite{ref:Whipple} are also considerably lower than
the earlier reports.
The limits presented here are a factor of
130 lower at $115\,$TeV, and a factor of 900 at $1175\,$TeV,
than the spectrum 
plotted in Figure~\ref{fig:OldCyg}.
Our results are also inconsistent with emission reported by
a smaller experiment using data 
taken during a time that overlapped our observations \cite{ref:Muraki}.

The limits presented here on transient emission from Cygnus X-3
are lower than, but in agreement with, those reported by
other air shower experiments.
There have been no compelling reports of transient emission of
gamma-rays from Cygnus X-3 over the period 1990-1995,
including during large radio flares from the source.
There was an observation of underground muons from the direction
of Cygnus X-3 during the January 1991 radio flare \cite{ref:Thomson}.
The reported flux for this observation was $7.5\times 10^{-10}$
muons cm$^{-2}$ sec$^{-1}$,
for muon energies above $0.7\,$TeV.
If the muons were produced in air showers by the interaction
of a hypothetical neutral particle from Cygnus X-3, we would expect
a typical neutral particle energy of $\sim 10\,$TeV 
\cite{ref:Gaisser2}.
Assuming that the particle spectrum continues to energies
detectable by CASA-MIA
(and conservatively using 
a soft spectrum comparable to the cosmic rays),
one derives an expected flux of $\sim 10^{-11}$
particles cm$^{-2}$ sec$^{-1}$ 
for energies above $115\,$TeV.
This flux is
a factor of 5 to 10 above the flux limits set by CASA-MIA on
the emission of any neutral particle during the January 1991 flare 
(Table~\ref{tab:flare_results}).

We have also shown that there is no evidence 
for $4.8\,$
periodic emission from Cygnus X-3.
This result is consistent with reports by other experiments
over the same period of time.
The limits on pulsed gamma-ray emission presented here for
the phase intervals of 0.2-0.3 and 0.6-0.7 (Table~\ref{tab:CygPhase})
are lower at $115\,$TeV, and considerably lower at $530\,$TeV,
than the fluxes predicted by a recent 
theoretical paper \cite{ref:Mitra}.

\subsection{Hercules X-1}
\label{subsec:hercules_comparison}
\begin{figure} 
\centerline{\ \psfig{figure=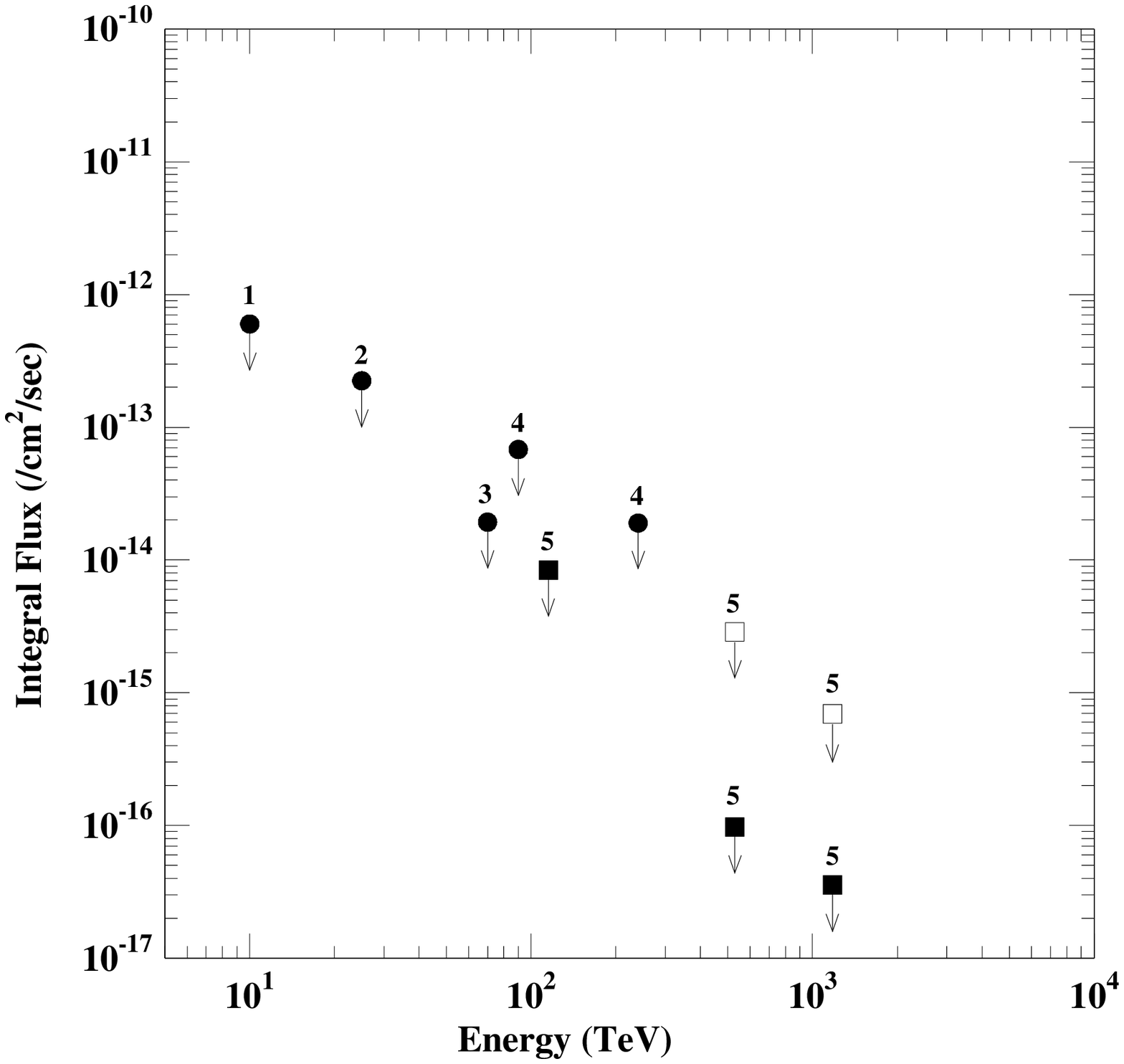,height=17.0cm}}
\caption{
Flux limits reported between 1990 and 1995 on the
steady emission of particles from Hercules X-1.
The squares (5) represent the results of this work.
Open squares indicate limits on the emission of any neutral
particle that creates air showers.
Filled
squares indicate limits on the emission of gamma-rays.
The circles (1-4) represent results from other experiments:
1. Tibet \protect\cite{ref:Amenomori},       
2. HEGRA \protect\cite{ref:Karle}, 
3. CYGNUS \protect\cite{ref:Alexandreas4},
and 4. EAS-TOP \protect\cite{ref:Aglietta}.}
\label{fig:NewHer}
\end{figure}

\begin{figure} 
\centerline{\ \psfig{figure=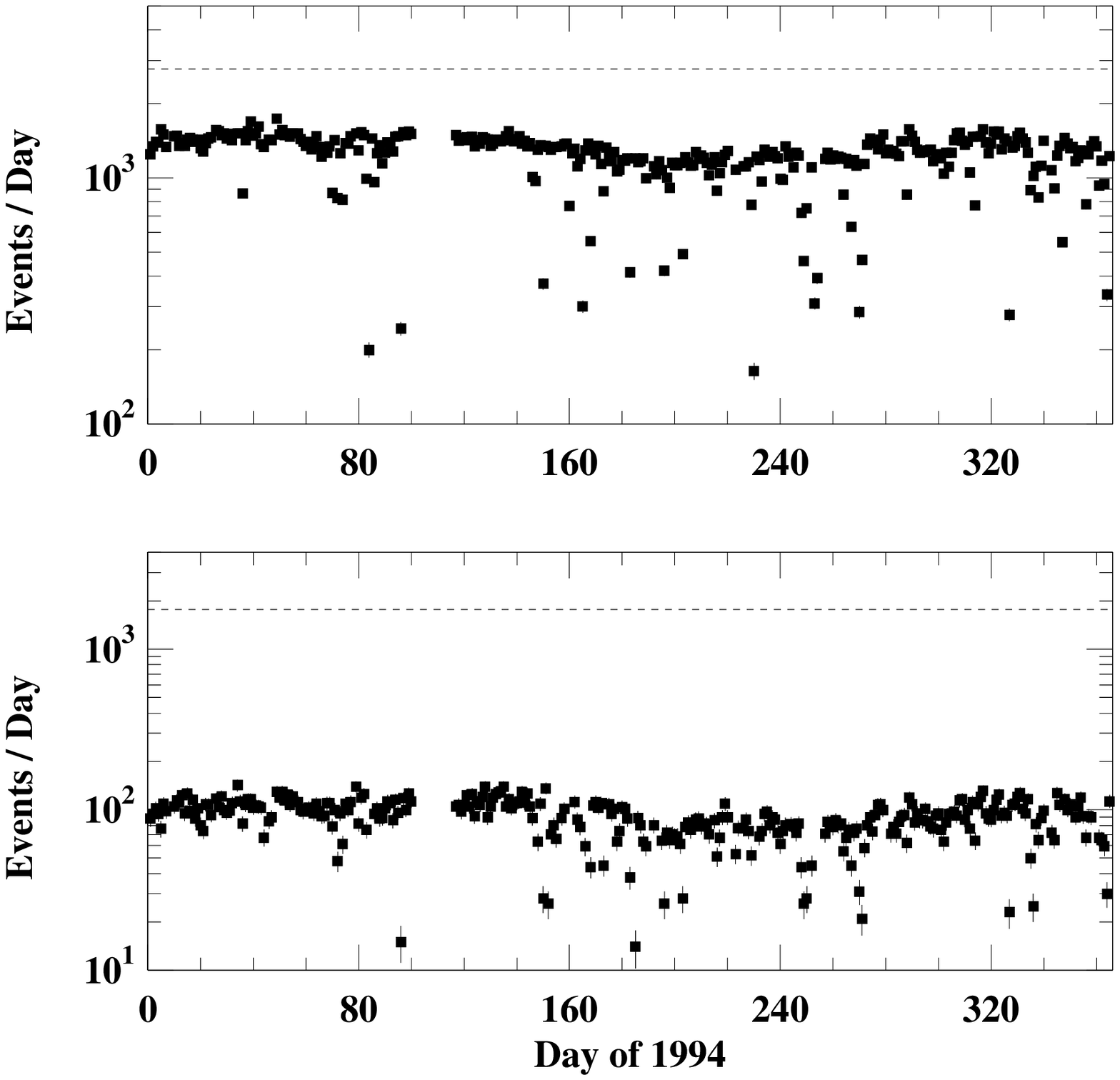,height=17.0cm}}
\caption{
Daily event totals in 1994 for showers
recorded by CASA-MIA from the
direction of Hercules X-1
for the all-data (top) and muon-poor (bottom)
samples.
The dashed lines indicate the expected numbers of events for
gamma-ray fluxes comparable to those seen 
in 1986 \protect\cite{ref:Dingus2}.
The gaps in the distribution indicate those
periods in which the experiment 
recorded no usable data from Hercules X-1.}
\label{fig:HerTran1994}
\end{figure}

The limits on steady emission of gamma-rays from Hercules X-1
presented here are in agreement with those from other
experiments, as shown in Figure~\ref{fig:NewHer}.
Gamma-ray emission from Hercules X-1 was typically seen by
earlier experiments as transient emission over short time
scales (e.g. the 1986 outbursts).
We have no evidence for such emission over the entire period 1990-1995.
In Figure~\ref{fig:HerTran1994}, we compare the 
daily event totals
observed by CASA-MIA from the direction of Hercules X-1 
to the total expected assuming the flux 
of an earlier reported outburst \cite{ref:Dingus2}.
Clearly, no evidence for
emission at even
much weaker levels than this outburst is seen during this time.
The flux reported in Ref.~\cite{ref:Dingus2} was $\sim 2\times 10^{-11}$
particles cm$^{-2}$ sec$^{-1}$ for minimum energies of $100\,$TeV.
This flux is about a factor of 45 larger than the typical
limits placed by CASA-MIA during the early part of operations
and about a factor of 80 larger than the typical daily gamma-ray
limits placed by the full CASA-MIA experiment (Table~\ref{tab:daily}).
Since we have no evidence for transient emission from Hercules X-1,
we choose not to carry out a periodicity analysis based on the
X-ray pulsar period of $1.24\,$sec.

\section{Conclusions}
\label{sec:conclusions}
We have carried out a high statistics search for ultra-high energy
neutral and gamma-ray particle emission from Cygnus X-3 and
Hercules X-1 between 1990 and 1995.
We have no evidence for steady or transient emission from
either source, and for Cygnus X-3, we have no evidence for
$4.8\,$hr
periodic emission or emission correlated with large radio
flares.
These results are in agreement with those from other experiments
operating during the same period of time, but are in stark
contrast to earlier (1975-1990) reports.

The apparent disappearance of Cygnus X-3 and Hercules X-1 from
the ultra-high energy gamma-ray sky can be interpreted in
two ways.
An optimistic view \cite{ref:Protheroe}
is that the earlier results indicated the presence of
ultra-high energy gamma-rays (or particles) from Cygnus X-3
and Hercules X-1, and that the sources, which are episodic on long times
scales, are now dormant.
A more pessimistic view is that the earlier reported detections were
largely, if not entirely, statistical fluctuations, and that no compelling
evidence exists for ultra-high energy gamma-rays from 
any astrophysical source.
We point out that an earlier all-sky survey using a portion of our data
sample indicates that the northern hemisphere does not contain
any steady
point sources of gamma-rays with fluxes comparable to those
reported from X-ray binaries in the 1980's
\cite{ref:McKay}.
We have presented an update on this analysis at a conference \cite{ref:Nitz}
which are consistent with the absence of bright 100 TeV gamma-ray
point sources.
We are in the process of completing a final all-sky survey on the
five year CASA-MIA data sample.

The pessimistic interpretation of the ultra-high energy point source
question, if correct,
highlights the difficulties in detecting gamma-rays from sources at
other (high) energies and in detecting neutrinos as well.
In addition, without compelling evidence for high energy particle
acceleration at point sources, the difficulties in explaining
the origins of cosmic rays above $10^{14}\,$eV remain.

%******************************************************************************
% END OF TEXT
%******************************************************************************

% Acknowledgements

\vspace{5mm}
\noindent {\bf Acknowledgements}
\vspace{3mm}

We acknowledge the assistance of the command and
staff of Dugway Proving Ground, and the University of Utah
Fly's Eye group.
Special thanks go to M. Cassidy.
We also wish to thank P.Burke, S. Golwala, M. Galli, J. He, H. Kim, L. Nelson,
M. Oonk, M. Pritchard, P. Rauske, K. Riley, and Z. Wells 
for assistance with data processing.
This work is supported by the U.S. 
National
Science Foundation and the U.S. Department of Energy.
JWC and RAO wish to acknowledge the support of
the W.W. Grainger Foundation.
RAO acknowledges additional support from the Alfred P. Sloan
Foundation.

\vspace{10mm}

\noindent $^*$ Present Address: Department of Physics, Massachusettts Institute
of Technology, Cambridge, MA 02139, USA.

\noindent $^\dag$ Present Address: Department of Physics and Astronomy,
Iowa State University, Ames, IA 50011, USA.

% Acknowledgements and references end here
% Figures

% Figures end here


\begin{thebibliography}{99}

\bibitem[1]{ref:Snowmass}
For a recent review of ground-based gamma-ray astronomy, see
R.C. Lamb, R.A. Ong, C.E. Covault, and D.A. Smith,
{\em Proc. of the 1994 Snowmass Summer Study,
Particle and Nuclear
Astrophysics and Cosmology in the Next 
Millennium}, 
ed. by E.W. Kolb and R.D. Peccei
(World Scientific, Singapore, 1995) 295.

\bibitem[2]{ref:Reviews}
For general reviews see:
T.C. Weekes, Phys. Rep. {\bf 160}, 1 (1988);
D.E. Nagle, T.K. Gaisser, and R.J. Protheroe,
Annu. Rev. Nucl. Part. Sci. {\bf 38}, 609 (1988);
T.C. Weekes, Space Sci. Rev. {\bf 59}, 315 (1992);
J.W. Cronin, K.G. Gibbs, and T.C. Weekes,
Annu. Rev. Nucl. Part. Sci. {\bf 43}, 883 (1993).

\bibitem[3]{ref:Protheroe}
R.J. Protheroe, Astrophys. J. Suppl. {\bf 90}, 883 (1994).

\bibitem[4]{ref:Chardin}
J.M. Bonnet-Bidaud and G. Chardin,
Phys. Rep. {\bf 170}, 325 (1988).

\bibitem[5]{ref:Giacconi}
R. Giacconi, P. Gorenstein, H. Gursky, and J.R. Walters,
Astrophys. J. {\bf 148}, L119 (1967).

\bibitem[6]{ref:Parsignault}
D.R. Parsignault, E. Schreier, J. Grindlay, and H. Gursky,
Astrophys. J. {\bf 209}, L73 (1976).

\bibitem[7]{ref:vanderKlis1}
M. van der Klis and J.M. Bonnet-Bidaud,
Astron. Astrophys. {\bf 95}, L5 (1981).

\bibitem[8]{ref:vanderKlis2}
M. van der Klis and J.M. Bonnet-Bidaud,
Astron. Astrophys. {\bf 214}, 203 (1989).

\bibitem[9]{ref:Kitamoto}
S. Kitamoto {\it et al.}, Publ. Astron. Soc. Japan {\bf 47},
233 (1995).

\bibitem[10]{ref:Gregory1}
P.C. Gregory {\it et al.}, Nature {\bf 239}, 440 (1972).

\bibitem[11]{ref:Johnston}
K.J. Johnston {\it et al.}, Astrophys. J. {\bf 309}, 707 (1986).

\bibitem[12]{ref:Waltman1}
E.B. Waltman {\it et al.}, Astronom. J. {\bf 108}, 179 (1994).

\bibitem[13]{ref:Waltman2}
E.B. Waltman {\it et al.}, Astronom. J. {\bf 110}, 290 (1995).

\bibitem[14]{ref:Lauque}
Robert Lauqu\'e, James Lequeux, and Nguyen-Quang-Rieu,
Nature Physical Science {\bf 239}, 119 (1972).

\bibitem[15]{ref:Lamb1}
R.C. Lamb {\it et al.}, Astrophys. J. {\bf 212}, L63 (1977).

\bibitem[16]{ref:Bennett}
K. Bennett {\it et al.}, Astron. Astrophys. {\bf 59}, 273 (1977).

\bibitem[17]{ref:Hermsen}
W. Hermsen {\it et al.}, Astron. Astrophys. {\bf 175}, 141 (1987).

\bibitem[18]{ref:Fichtel}
C.E. Fichtel {\it et al.}, Astrophys. J. {\bf 319}, 362 (1987).

\bibitem[19]{ref:Li}
Ti-Pei Li and Mei Wu, Astrophys. J. {\bf 346}, 391 (1989).

\bibitem[20]{ref:Michelson}
P.F. Michelson {\it et al.}, Astrophys. J. {\bf 401}, 724 (1992).

\bibitem[21]{ref:Neshpor}
Yu.I. Neshpor {\it et al.}, Astrophys. Space Sci. {\bf 61}, 349 (1979).

\bibitem[22]{ref:Danaher}
S. Danaher, D.J. Fegan, N.A. Porter, and T.C. Weekes,
Nature {\bf 289}, 568 (1981).

\bibitem[23]{ref:Lamb2}
R.C. Lamb, C.P. Godfrey, W.A. Wheaton, and T. Tumer,
Nature {\bf 296}, 543 (1982).

\bibitem[24]{ref:Dowthwaite1}
J.C. Dowthwaite {\it et al.}, Astron. Astrophys. {\bf 126}, 1 (1983).

\bibitem[25]{ref:Cawley1}
M.F. Cawley {\it et al.}, Astrophys. J. {\bf 296}, 185 (1985).

\bibitem[26]{ref:Chadwick}
P.M. Chadwick {\it et al.}, Nature {\bf 318}, 642 (1985).

\bibitem[27]{ref:Bhat}
C.L. Bhat, M.L. Sapru, and H. Razdan, Astrophys. J. {\bf 306},
587 (1986).

\bibitem[28]{ref:Brazier}
K.T.S. Brazier {\it et al.}, Astrophys. J. {\bf 350}, 745 (1990).

\bibitem[29]{ref:Gregory2}
A.A. Gregory {\it et al.}, Astron. Astrophys. {\bf 237}, L5 (1990).

\bibitem[30]{ref:Samorski}
M. Samorski and W. Stamm, Astrophys. J. {\bf 268}, L17 (1983).

\bibitem[31]{ref:Lloyd}
J. Lloyd-Evans {\it et al.}, Nature {\bf 305}, 784 (1983).

\bibitem[32]{ref:Kifune}
T. Kifune {\it et al.}, Astrophys. J. {\bf 301}, 230 (1986).

\bibitem[33]{ref:Alexeenko}
V.V. Alexeenko {\it et al.}, Il Nuovo Cimento {\bf 10C}, 151 (1987).

\bibitem[34]{ref:Baltrusaitis1}
R.M. Baltrusaitis {\it et al.}, Astrophys. J. {\bf 323}, 685 (1987).

\bibitem[35]{ref:Tonwar1}
S.C. Tonwar, N.V. Gopalakrishnan, M.R. Rajeev, and
B.V. Sreekantan, Astrophys. J. {\bf 330}, L107 (1988).

\bibitem[36]{ref:Morello}
C. Morello, L. Periale, P. Vallania, and G. Navarra,
Il Nuovo Cimento {\bf 13C}, 453 (1990).

\bibitem[37]{ref:Cassiday1}
G.L. Cassiday {\it et al.}, Phys. Rev. Lett. {\bf 62}, 383 (1989).

\bibitem[38]{ref:Teshima}
M. Teshima {\it et al.}, Phys. Rev. Lett. {\bf 64}, 1628 (1990).

\bibitem[39]{ref:Lawrence}
M.A. Lawrence, D.C. Prosser, and A.A. Watson,
Phys. Rev. Lett. {\bf 63}, 1121 (1989).

\bibitem[40]{ref:Cawley2}
M.F. Cawley and T.C. Weekes,
Astron. Astrophys. {\bf 133}, 80 (1984).

\bibitem[41]{ref:Dingus1}
B.L. Dingus {\it et al.}, Phys. Rev. Lett. {\bf 60}, 1785 (1988).

\bibitem[42]{ref:Fegan}
D.J. Fegan {\it et al.}, 
Astron. Astrophys. {\bf 211}, L1 (1989).

\bibitem[43]{ref:Cassiday2}
G.L. Cassiday {\it et al.}, Phys. Rev. Lett. {\bf 63}, 2329 (1989).

\bibitem[44]{ref:Alexandreas1}
D.E. Alexandreas {\it et al.}, Phys. Rev. Lett. {\bf 64}, 2973 (1990).

\bibitem[45]{ref:Ciampa}
D. Ciampa {\it et al.}, Phys. Rev. {\bf D42}, 281 (1990).

\bibitem[46]{ref:Cronin}
J.W. Cronin {\it et al.}, Phys. Rev. {\bf D45}, 4385 (1992).

\bibitem[47]{ref:Muraki}
Y. Muraki {\it et al.}, Astrophys. J. {\bf 373}, 657 (1991).

\bibitem[48]{ref:Bowden}
C.C.G. Bowden {\it et al.}, J. Phys. G {\bf 18}, 413 (1992).

\bibitem[49]{ref:Tonwar2}
S.C. Tonwar {\it et al.}, Astrophys. J. {\bf 390}, 273 (1992).

\bibitem[50]{ref:Tananbaum}
H. Tananbaum {\it et al.}, Astrophys. J. {\bf 174}, L143 (1972).

\bibitem[51]{ref:Jones}
Christine A. Jones, William Forman, and William Liller,
Astrophys. J. {\bf 182}, L109 (1973).

\bibitem[52]{ref:Forman}
William Forman, Christine A. Jones, and William Liller,
Astrophys. J. {\bf 177}, L103 (1972).

\bibitem[53]{ref:Deeter}
J.E. Deeter, P.E. Boynton, and S.H. Pravdo,
Astrophys. J. {\bf 247}, 1003 (1981).

\bibitem[54]{ref:Dowthwaite2}
J.C. Dowthwaite {\it et al.}, 
Nature {\bf 309}, 691 (1984).

\bibitem[55]{ref:Baltrusaitis2}
R.M. Baltrusaitis {\it et al.}, Astrophys. J. {\bf 293}, 
L69, (1985).

\bibitem[56]{ref:Gorham1}
P.W. Gorham {\it et al.}, Astrophys. J. {\bf 308},
L11 (1986).

\bibitem[57]{ref:Gorham2}
P.W. Gorham {\it et al.}, Astrophys. J. {\bf 309},
114 (1986).

\bibitem[58]{ref:Resvanis}
L.K. Resvanis {\it et al.}, Astrophys. J. {\bf 328},
L9 (1988).

\bibitem[59]{ref:Lamb3}
R.C. Lamb {\it et al.}, Astrophys. J. {\bf 328},
L13 (1988).

\bibitem[60]{ref:Dingus2}
B.L. Dingus {\it et al.}, Phys. Rev. Lett. {\bf 61},
1906 (1988).

\bibitem[61]{ref:Vishwanath}
P.R. Vishwanath, P.N. Bhat, P.V. Ramanamurthy, and B.V. Sreekantan,
Astrophys. J. {\bf 342}, 489 (1989).

\bibitem[62]{ref:Gupta}
S.K. Gupta {\it et al.}, Astrophys. J. {\bf 354},
L13 (1990).

\bibitem[63]{ref:Reynolds}
P.T. Reynolds {\it et al.}, Astrophys. J. {\bf 382},
640 (1991).

\bibitem[64]{ref:Alexandreas2}
D.E. Alexandreas {\it et al.}, Astrophys. J. {\bf 383},
L53 (1991).

\bibitem[65]{ref:Cheng1}
K.S. Cheng, C. Ho, and M. Ruderman, Astrophys. J. {\bf 300},
522 (1986).

\bibitem[66]{ref:Chanmugam}
G. Chanmugam and K. Brecher,
Nature {\bf 313}, 767 (1985).

\bibitem[67]{ref:Vestrand}
W.Thomas Vestrand and David Eichler,
Astrophys. J. {\bf 261}, 251 (1982).

\bibitem[68]{ref:Eichler1}
David Eichler and W.Thomas Vestrand,
Nature {\bf 307}, 613 (1984).

\bibitem[69]{ref:Kazanas}
Demosthenes Kazanas and Donald C. Ellison,
Nature {\bf 319}, 380 (1986).

\bibitem[70]{ref:Wdowczyk}
J. Wdowczyk and A.W. Wolfendale,
Nature {\bf 305}, 609 (1983).

\bibitem[71]{ref:Hillas}
A.M. Hillas, Nature {\bf 312}, 50 (1984).

\bibitem[72]{ref:Eichler2}
David Eichler and W.Thomas Vestrand,
Nature {\bf 318}, 345 (1985).

\bibitem[73]{ref:Gorham3}
Peter W. Gorham and John G. Learned,
Nature {\bf 323}, 422 (1986).

\bibitem[74]{ref:Cheng2}
K.S. Cheng and Malvin Ruderman,
Astrophys. J. {\bf 337}, L77 (1989).

\bibitem[75]{ref:Slane}
P. Slane and W.F. Fry,
Astrophys. J. {\bf 342}, 1129 (1989).

\bibitem[76]{ref:NIMPaper}
A. Borione {\it et al.}, Nucl. Inst. Meth. {\bf A346}, 329 (1994).

\bibitem[77]{ref:Moon}
A. Borione {\it et al.}, Phys. Rev. {\bf D49}, 1171 (1994).

\bibitem[78]{ref:HERA}
M. Derrick {\it et al.}, Phys. Lett. {\bf B293}, 465 (1992).
T. Ahmed {\it et al.}, Phys. Lett. {\bf B299}, 469 (1993).

\bibitem[79]{ref:Chatelet}
E. Chatelet {\it et al.}, J. Phys. G. 
{\bf 16}, 317 (1990).

\bibitem[80]{ref:Halzen}
T.K. Gaisser {\it et al.}, Phys. Rev. {\bf D43}, 314 (1991).

\bibitem[81]{ref:Alexandreas3}
D.E. Alexandreas {\it et al.}, Nucl. Inst. Meth.
{\bf A328}, 570 (1993).

\bibitem[82]{ref:Nagano}
M. Nagano {\it et al.}, J. Phys. G.
{\bf 18}, 423 (1992).

\bibitem[83]{ref:McKay}
T.A. McKay {\it et al.}, Astrophys. J. {\bf 417}, 742 (1993).

\bibitem[84]{ref:Newport}
A. Borione {\it et al.}, 
``A Search for Ultrahigh Energy Gamma-Ray Emission from the Crab
Nebula and Pulsar,'' EFI 96-26 (August 1996), submitted to
The Astrophysical Journal.

\bibitem[85]{ref:Gaisser1}
T.K. Gaisser {\it et al.}, Phys. Rev. Lett. {\bf 62}, 1425 (1989).

\bibitem[86]{ref:Asakimori}
K. Asakimori {\it et al.}, Proc. 23rd Int. Cosmic Ray Conf. (Calgary),
ed. D.A. Leahy, R.B. Hicks, and D. Venkatesan
(World Scientific Publishing, Singapore),
{\bf 2}, 21 (1994).

\bibitem[87]{ref:Swordy}
S.P. Swordy, Proc. 23rd Int. Cosmic Ray Conf. (Calgary),
ed. D.A. Leahy, R.B. Hicks, and D. Venkatesan
(World Scientific Publishing, Singapore),
Rapporteur Volume, 243 (1994).

\bibitem[88]{ref:LiMa}
T. Li and Y. Ma, Astrophys. J. {\bf 272}, 317 (1983).

\bibitem[89]{ref:Helene}
O. Helene, Nucl. Inst. Meth. {\bf 212}, 319 (1983).

\bibitem[90]{ref:PDG}
Review of Particle Properties, Phys. Rev. {\bf D50},
1280 (1994).

\bibitem[91]{ref:Standish}
E.M. Standish,
Astron. Astrophys. {\bf 233}, 252 (1990).

\bibitem[92]{ref:Amenomori}
M. Amenomori {\it et al.}, Phys. Rev. Lett.
{\bf 69}, 2468 (1992).

\bibitem[93]{ref:Alexandreas4}
D.E. Alexandreas {\it et al.},
Astrophys. J. {\bf 405}, 353 (1993); updated in
D.E. Alexandreas {\it et al.}, Proc. XXIII Int. Cosmic
Ray Conf. (Calgary),
ed. D.A. Leahy, R.B. Hicks, and D. Venkatesan
(World Scientific Publishing, Singapore),
{\bf 1}, 353 (1994).

\bibitem[94]{ref:Aglietta}
M. Aglietta {\it et al.},
Astropart. Phys. {\bf 3}, 1 (1995).

\bibitem[95]{ref:Karle}
A. Karle {\it et al.}
Astropart. Phys. {\bf 4}, 1 (1996).

\bibitem[96]{ref:Whipple}
K.S. O'Flaherty {\it et al.},
Astrophys. J. {\bf. 396}, 674 (1992).

\bibitem[97]{ref:Thomson}
M.A. Thomson {\it et al.}
Phys. Lett. {\bf B269}, 220 (1991).

\bibitem[98]{ref:Gaisser2}
Thomas K. Gaisser, {\it Cosmic Rays and Particle Physics},
Cambridge University Press (1990).

\bibitem[99]{ref:Mitra}
Abhas Mitra, Astrophys. J. {\bf 425}, 782 (1994).

\bibitem[100]{ref:Nitz}
A. Borione {\it et al.}, 
Proc. 23rd Int. Cosmic Ray Conf. (Calgary),
ed. D.A. Leahy, R.B. Hicks, and D. Venkatesan
(World Scientific Publishing, Singapore),
{\bf 1}, 357 (1994).

\end{thebibliography}
\end{document}